%% file: nature.tex
\documentclass{natureprintstyle}
\usepackage{epsfig}
\usepackage{color}
\usepackage{bm}
\usepackage{graphicx}
\usepackage{longtable}
\usepackage{amssymb}
\usepackage{rotating}
\def\arcsec{\hbox{$^{\prime\prime}$}}
\newcommand{\sol}{$_{\odot}$}

\newcommand{\apj}{Astrophys. J.}

\newcommand{\apjs}{Astrophys. J. Supp.}
\newcommand{\araa}{Annu. Rev. Astron. Astrophys.}
\newcommand{\mnras}{Mon. Not. R. Astron. Soc.}
\newcommand{\apjl}{Astrophys. J. Let.}
\newcommand{\aap}{Astron. Astrophys.}
\newcommand{\aj}{Astron. J.}
\newcommand{\nat}{Nature}

\newcommand{\aaps}{A\&AS}
\newcommand{\gt}{$>$}

\title{A Rapidly Star-forming Galaxy 700 Million Years After the Big Bang at z=7.51}

\author{S.~L.~Finkelstein$^{1}$, C.~Papovich$^2$, M.~Dickinson$^3$,
  M.~Song$^1$, V.~Tilvi$^2$, A.~M.~Koekemoer$^4$, K.~D.~Finkelstein$^1$,
  B.~Mobasher$^5$, H.~C.~Ferguson$^4$, M.~Giavalisco$^6$,
  N.~Reddy$^5$, M.~L.~N.~Ashby$^7$, A.~Dekel$^{8}$, G.~G.~Fazio$^7$,
  A.~Fontana$^{9}$, N.~A.~Grogin$^4$,
  J.-S.~Huang$^{7}$, D.~Kocevski$^{10}$, M.~Rafelski$^{11}$, B.\ J.\
  Weiner$^{12}$ \& S.~P.~Willner$^{7}$}

\begin{document}

\maketitle

\let\thefootnote\relax\footnote{
\begin{affiliations}
\item The University of Texas at Austin, 2515 Speedway, Stop C1400,
  Austin, TX 78712, USA
\item George P. and Cynthia Woods Mitchell Institute for Fundamental
  Physics and Astronomy, Texas A\&M University, 4242 TAMU, College Station, TX 78743, USA
\item National Optical Astronomy Observatory, Tucson, AZ 85719, USA
\item Space Telescope Science Institute, 3700 San Martin Dr., Baltimore, MD, 21218, USA
\item University of California, Riverside, CA 92521, USA
\item University of Massachusetts, Amherst, MA 01003, USA
\item Harvard-Smithsonian Center for Astrophysics, 60 Garden St.,
  Cambridge, MA 02138, USA
\item Racah Institute of Physics, The Hebrew University, Jerusalem 91904, Israel
\item INAF-Osservatorio di Roma,II-00040, Monteporzio, Italy
\item University of Kentucky, Lexington, KY, 40506, USA
\item Infrared Processing and Analysis Center, MS 100-22, Caltech,
  Pasadena, CA 91125
\item Steward Observatory, University of Arizona, 933 N. Cherry Ave, Tucson, AZ 85721
\end{affiliations}
}

\vspace{-3.5mm}
\begin{abstract}
Out of several dozen z $>$ 7 candidate galaxies observed
spectroscopically, only five have been confirmed via
Lyman$\boldsymbol{\alpha}$ emission, at
z=7.008, 7.045, 7.109, 7.213 and 7.215.\cite{vanzella11,schenker12,ono12,shibuya13}
The small fraction of confirmed galaxies may indicate that the neutral fraction in the
intergalactic medium (IGM) rises quickly at $z >$
6.5, as Lyman$\boldsymbol{\alpha}$ is resonantly scattered by neutral gas.\cite{kashikawa06,iye06,ouchi10,pentericci11,ono12}
However, the small samples and limited depth of
previous observations makes these conclusions tentative.
Here we report the results of a deep near-infrared spectroscopic
survey of 43 $z >$ 6.5 galaxies.  We detect
only a single galaxy, confirming that some process is making
Lyman$\boldsymbol{\alpha}$ difficult to detect.
The detected emission line at 1.0343\,$\boldsymbol{\mu}$m is likely to be
Lyman$\boldsymbol{\alpha}$ emission, placing
this galaxy at a redshift z = 7.51, an epoch 700 million years
after the Big Bang.  This galaxy's colors are consistent with
significant metal content, implying that galaxies become
enriched rapidly.  We measure a surprisingly high star formation rate 
of 330 M\sol\ yr$^{-1}$, more than a factor of 100 greater than seen
in the Milky Way.  Such a galaxy is unexpected in
a survey of our size\cite{smit12}, suggesting that the early universe may harbor more
intense sites of star-formation than expected.
\end{abstract}

We obtained near-infrared (near-IR) spectroscopy of galaxies originally discovered in the
Cosmic Assembly Near-infrared Deep
Extragalactic Legacy Survey (CANDELS)\cite{grogin11,koekemoer11}  with
the newly-commissioned near-infrared
spectrograph MOSFIRE\cite{mclean12} on the Keck {\sc{i}} 10 meter telescope.
From a parent sample
of over 100 galaxy candidates at $z >$ 7 selected via
their {\it HST} colors through the photometric redshift
technique\cite{finkelstein10a,mclure10,finkelstein12a,mclure13}, we observed 43 candidate high-redshift galaxies over two MOSFIRE
pointings with exposure times of 5.6 and 4.5 hr, respectively.  Our
observations covered Ly$\alpha$ emission at redshifts of 7.0 -- 8.2.
We visually inspected the reduced data at the expected slit positions
for our 43 observed sources and found plausible
emission lines in eight objects, with only one line detected at $>$5$\sigma$
significance.  The detected emission line is at a
wavelength of 1.0343$\mu$m with an integrated signal-to-noise (S/N) of
7.8 (Figure 1) and comes from the object designated z8\_GND\_5296 in
our sample (RA$=$12:36:37.90; Dec$=$62:18:08.5 J2000).  
Based on the arguments outlined below (and discussed extensively in
the supplementary material), we identify this line as the Ly$\alpha$
transition of hydrogen at a line-peak redshift of $z =$ 7.5078 $\pm$ 0.0004;
consistent with our photometric redshift 68\% confidence range of 7.5
$< z <$ 7.9 for z8\_GND\_5296.

\begin{figure*}
\centerline{
\includegraphics[width=0.7\textwidth]{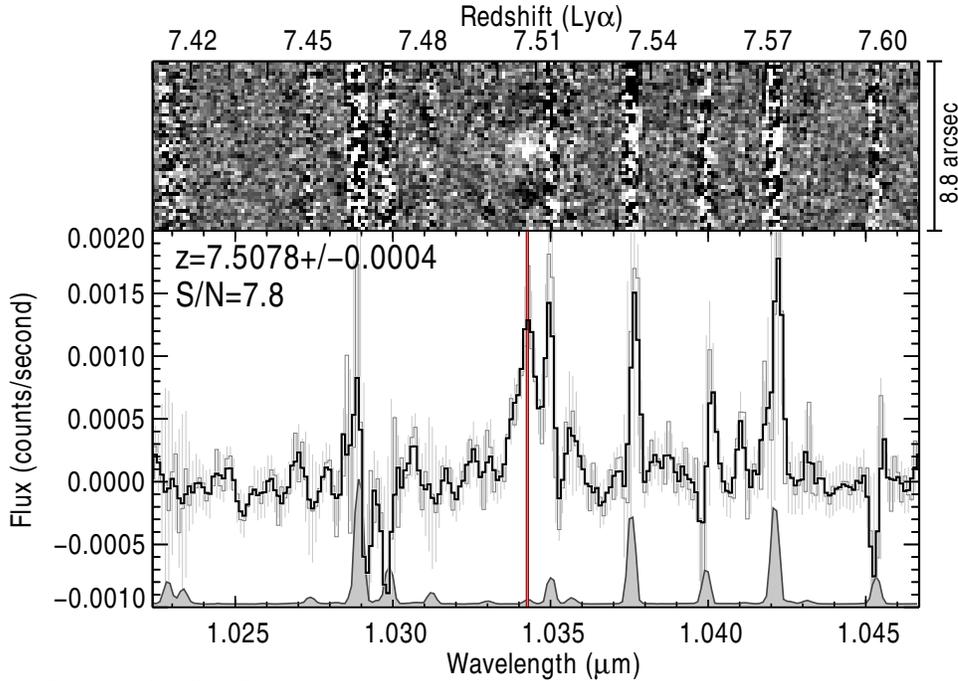}
}
\vspace{-4mm}
\caption{\textbf{MOSFIRE Spectrum of z8\_GND\_5296}. The observed near-IR spectrum of
  the galaxy z8\_GND\_5296.  The top panel shows the reduced
  two-dimensional spectrum, and an emission line is clearly seen as a
  positive signal in the center, with the negative signals above and below a result of our dithering pattern in the
  spatial direction along the slit; this is a
  pattern only exhibited for real objects.  The bottom panel shows our
extracted one-dimensional spectrum (smoothed to the spectral
resolution in black; un-smoothed in gray).  The sky
spectrum is shown as the filled gray curve with the scale reduced
greatly compared to that of the data. 
The line is clearly detected in separate reductions of the first and
second halves of the data with S/N of 6.4 and 5.2, respectively.  The
line has a full-width at half-maximum (FWHM) of 7.7 \AA\ and is
clearly resolved compared to nearby sky emission lines, which have
FWHM $=$ 2.7 \AA.  The red
line denotes the peak flux of the detected emission line, which corresponds to a
redshifted Ly$\alpha$ line at $z =$ 7.51.   All other strongly positive or negative
features are subtraction residuals due to strong
night sky emission.  Although the line appears symmetric, there is a
sky line residual just to the red of our detected emission line, which makes a
measurement of our line's asymmetry difficult.}
\vspace{-4mm}
\end{figure*}

\begin{figure*}
\centerline{
\includegraphics[width=0.75\textwidth]{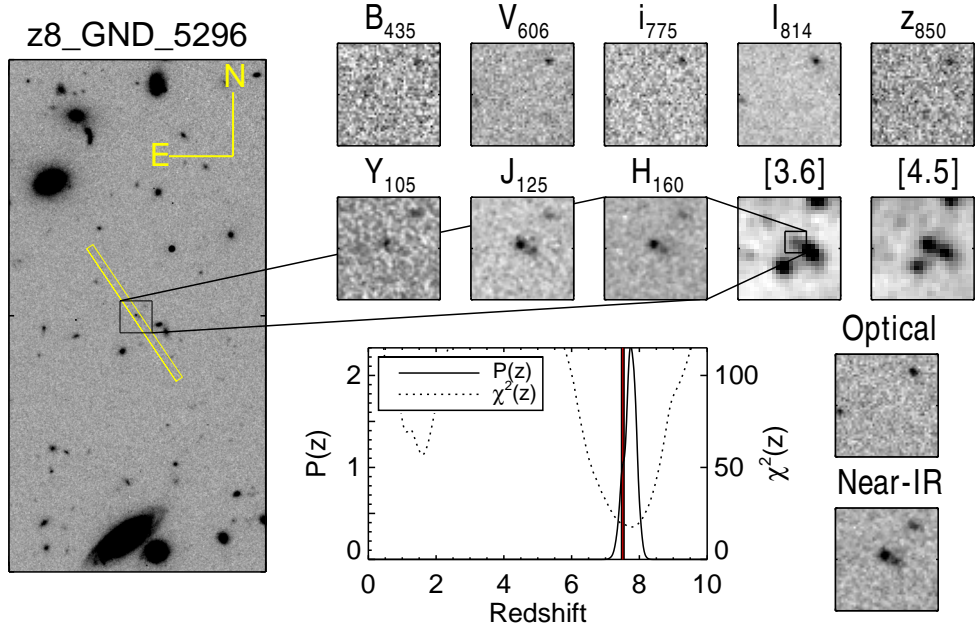}
}
\vspace{-4mm}
\caption{\textbf{Images of z8\_GND\_5296}. a) A portion of
  the CANDELS/GOODS-N field, shown in the F160W filter (centered at 1.6$\mu$m), around
z8\_GND\_5296.  CANDELS provides the largest
survey volume in the distant universe
deep enough to find $z >$ 7 galaxies.  The 15\arcsec\ $\times$ 0.7\arcsec\ slit is shown as
the yellow rectangle.  b) Cutouts around z8\_GND\_5296; the GOODS and
CANDELS {\it HST} images are 3\arcsec\ on a
side, while the S-CANDELS {\it Spitzer}/IRAC 3.6 and 4.5~$\mu$m images
are 15\arcsec on a side.  We also show mean stacks of the five optical
bands and the three near-IR bands, the latter of which shows that this
galaxy appears to have a clumpy morphology.  This galaxy is not detected in any
optical band, even when stacked together, which is strongly suggestive of a redshift greater than 7.  The
IRAC bands show a faint detection at 3.6~$\mu$m and a strong
detection at 4.5~$\mu$m.  This signature is expected if strong
[O\,{\sc iii}] emission is present in the 4.5~$\mu$m band, which would
be the case for a strongly star-forming galaxy at $z \sim$ 7.5 with
sub-Solar (though still significant) metal content
(0.2--0.4 Z\sol).  c) The results of our photometric redshift analysis
placing z8\_GND\_5296 at 7.3 $< z <$ 8.1 at 95\% confidence, which
encompasses our measured spectroscopic redshift (denoted by the vertical
line).  We show both the probability distribution function as well as
the values of $\chi^2$ at each redshift from the photometric redshift analysis; though a low-redshift
solution is possible, it is strongly disfavored, with the
high-redshift solution being $\sim$7$\times$10$^{9}$ more probable.}
\vspace{-4mm}
\end{figure*}

\begin{figure*}
\centerline{
\includegraphics[width=0.5\textwidth]{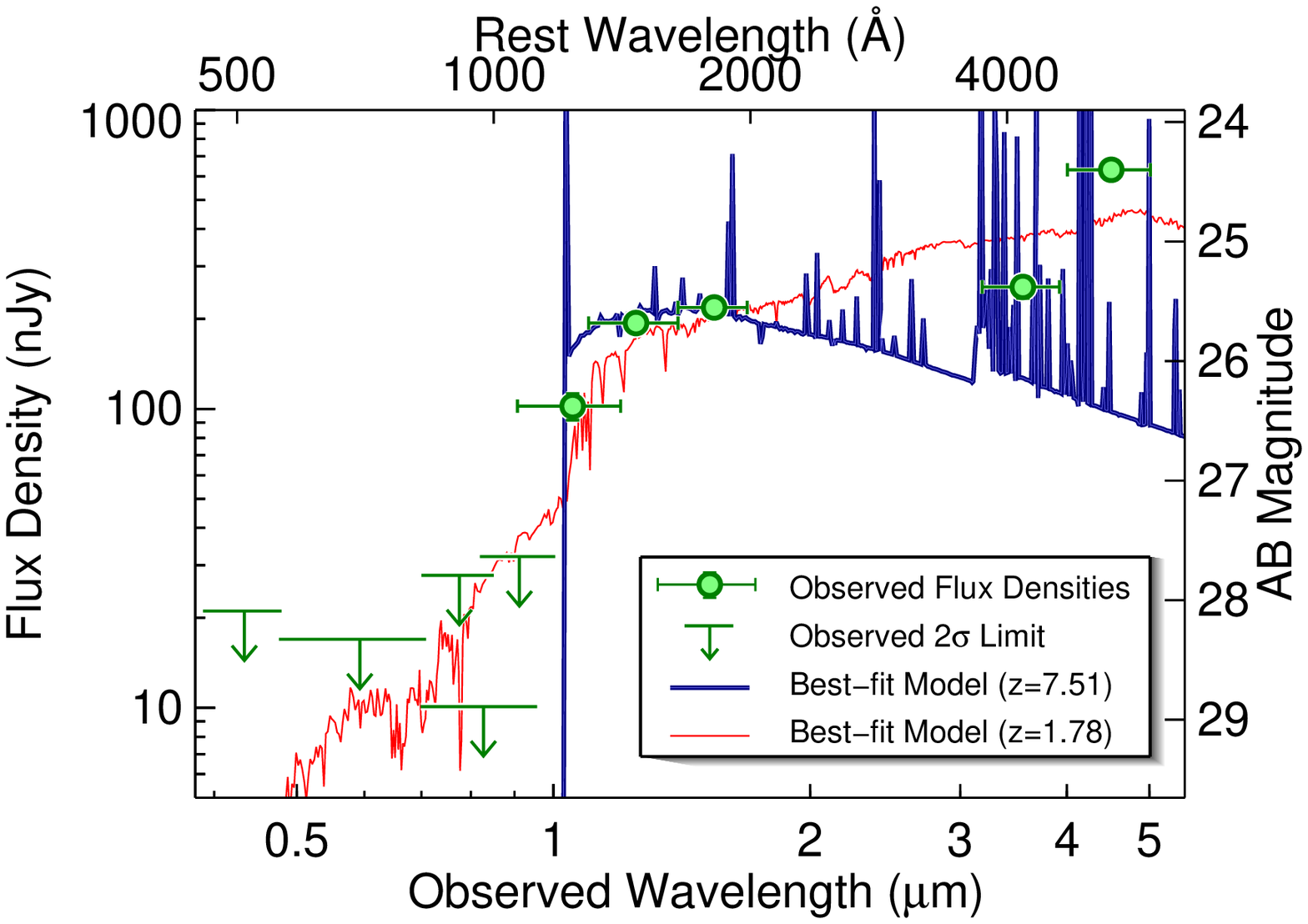}
\includegraphics[width=0.5\textwidth]{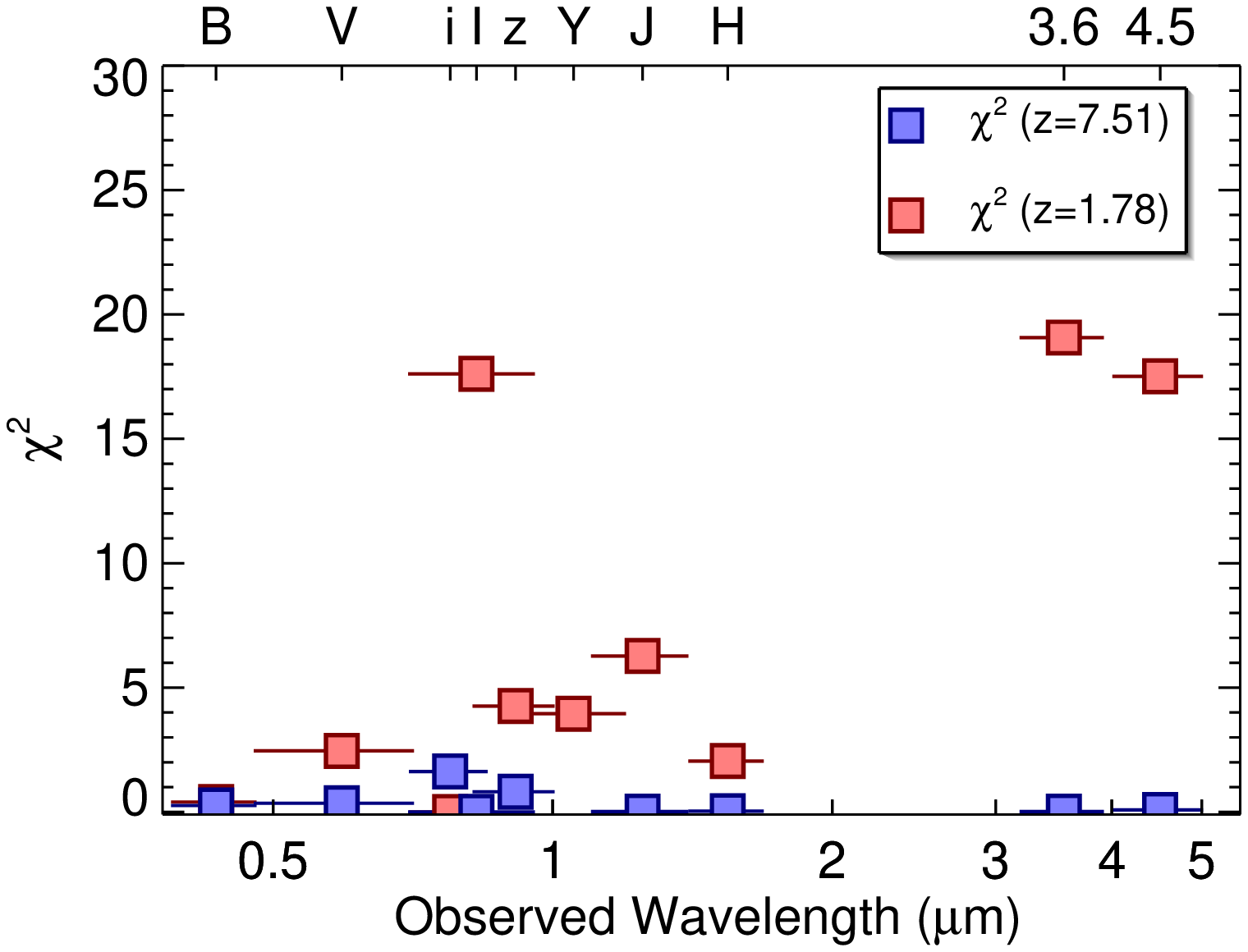}
}
\vspace{-5mm}
\caption{\textbf{Spectral Energy Distribution Fitting of
    z8\_GND\_5296}. a) The results of fitting stellar
  population models to the observed photometry of z8\_GND\_5296.  The
  best-fit model for $z=$7.51 (if the
  detected emission line is Ly$\alpha$) is shown by the blue spectrum, while the
alternate redshift of $z=1.78$ (if the line is
[O\,{\sc ii}]) is shown by the red spectrum.  The vertical error bars show the 1$\sigma$ flux
errors, while the horizontal error bars (in both panels) denote the
bandpass FWHM covered
by the filter.  b) The measured $\chi^2$ for
each band for the best-fit model at each redshift.  The lack of detectable
optical flux, particularly in the deep F814W image, as well as
the extremely red IRAC color, strongly favor the high-redshift
solution (reduced $\chi^2[z=7.51] =$ 0.8 versus $\chi^2[z=1.78] =$
14.7).  Additionally, the low-redshift model exhibits no
star-formation, thus this stellar population should not have detectable
[O\,{\sc ii}] emission.
The best-fitting high-redshift model shows that this galaxy has a 
stellar mass of about 10$^9$ M\sol, with a 10-Myr-averaged
star-formation rate (SFR) of $\sim$330 M\sol\ yr$^{-1}$ (68\% C.L. 320 --
1040 M\sol\ yr$^{-1}$).  The large SFR may be responsible for
the ability of Ly$\alpha$ to escape this galaxy.}
\vspace{-4mm}
\end{figure*}

 As expected for a galaxy at $z=$ 7.51, z8\_GND\_5296 is completely
 undetected in the {\it HST} optical bands,
including an extremely deep 0.8 $\mu$m image (Figure 2).  The galaxy is bright in the 
{\it HST} near-IR bands, becoming brighter with increasing wavelength,
implying that the Lyman break lies near 1 $\mu$m and that the galaxy
has a moderately red rest-frame UV color.  The galaxy is well-detected
 in both {\it Spitzer}/IRAC bands and is much brighter at 4.5 $\mu$m than at
3.6 $\mu$m.  The strong break at observed 1 $\mu$m restricts the observed emission line to be
either Ly$\alpha$ at $z =$ 7.51 (near the Lyman break) or
[O\,{\sc ii}] $\lambda\lambda$3726,3729 (a doublet) at $z =$ 1.78
(near the rest-frame Balmer/4000 \AA\ break).
We investigated these two possibilities by 
comparing our observed photometry to a suite of stellar population
models at both redshifts (Figure 3).  
A much better fit to the data is obtained when using models at $z =$
7.51 than at $z =$ 1.78, supporting our
identification of the emission line as Ly$\alpha$.  Specifically, the
model at $z =$ 1.78 would result in $>$4$\sigma$ significant flux in
the 0.8 $\mu$m image as well as a near-zero IRAC color, neither of
which is seen.
Additionally, the $z =$ 1.78 model
requires an aging stellar population with no active star formation, which would have insignificant
[O\,{\sc ii}] emission.  
In the Supplementary Information, we discuss a number of tests 
performed to discern between the Ly$\alpha$ and [O\.{\sc ii}]
hypotheses.  In summary, although we
cannot robustly measure the line asymmetry due to the nearby sky
residual, the SED fitting results and the lack of a detected
second line in the [O\,{\sc ii}] doublet lead us to conclude that the
detected emission
line is Ly$\alpha$ at $z =$ 7.51.

This galaxy is very bright in the rest-frame UV and optical, with an apparent magnitude of
m$_{F160W} =$ 25.6 and a derived stellar mass of
1.0$^{+0.2}_{-0.1}$ $\times$ 10$^{9}$ M\sol. The blue $H-[3.6]$ 
color suggests that the moderately red UV color ($J-H =$ 0.1 mag) is due to dust
attenuation rather than the intrinsic red color of an old stellar
population.  The presence of dust extinction leads to a higher
inferred UV luminosity.  To derive the intrinsic star-formation rate
(SFR) for this galaxy, we measured a time-averaged SFR from the
best-fitting stellar population models to find SFR
$=$ 330$^{+710}_{-10}$ M\sol\ yr$^{-1}$.
The very red $[3.6]-[4.5]$ color at $z =$ 7.51 can only be due to strong
[O\,{\sc iii}] $\lambda$5007 line emission in the 4.5~$\mu$m band; indeed, the SED
fitting implies an [O\,{\sc iii}] $\lambda$5007 rest-frame equivalent
width (EW) of 560 -- 640 \AA\ (68\%
confidence), with a line flux of 5.3 $\times$ 10$^{-17}$ erg s$^{-1}$
cm$^{-2}$.  This very high [O\,{\sc iii}] EW 
constrains the abundance of metals in this galaxy, as highly enriched
stars do not produce hard enough ionizing spectra, while
very low metallicity systems do not have enough oxygen to produce
strong emission lines.  Of the metallicities available in our models
(0.02, 0.2, 0.4 and 1.0 $\times$ solar), only models with a metal
abundance of about 20--40\% of Solar have [O\,{\sc
  iii}] EWs $>$300 \AA.  Thus, even at
such early times, it was possible to form a moderately chemically
enriched galaxy.  
However, due to discreteness of the
model metallicities, further analysis is needed to make more
quantitative conclusions on the metallicity, particularly on the lower
limit.  We note that at $z =$ 7.51 [O\,{\sc ii}] is in the 3.6$\mu$m
band, but it is predicted to be about 5$\times$ fainter than
[O\,{\sc iii}] and thus does not significantly affect the 3.6$\mu$m band.

This galaxy is forming stars at a very high rate,
with a ``mass-doubling'' time of at most 4 Myr.  
The most recent estimates \cite{stark13} at $z \approx$ 7 find
that galaxies with stellar masses of 5 $\times$10$^{9}$ M\sol\ typically
have specific SFRs (sSFR $=$ SFR divided by stellar mass)
$\sim$10$^{-8}$ yr$^{-1}$.  This galaxy is a factor of five less
massive, yet its sSFR is a factor of 30 greater at 3 $\times$
10$^{-7}$ yr$^{-1}$, implying that z8\_GND\_5296 is undergoing a
significant starburst.
Additionally, estimates of the SFR functions\cite{smit12}
show that a typical galaxy at $z \sim$ 7 has a SFR $=$ 10 M\sol\
yr$^{-1}$; the measured SFR of z8\_GND\_5296 is a factor of $>$30$\times$ greater.
If this SFR function is accurate, the expected
space density per co-moving Mpc$^{3}$ for this galaxy would be $\ll$
10$^{-5}$.  The implied rarity of this galaxy could imply that it is
the progenitor of some of the most massive systems in the high-redshift universe. 
However, the $z = 7.213$ galaxy GN-108036\cite{ono12}, also in GOODS-North, also
has an implied SFR $>$ 100 M\sol\ yr$^{-1}$.   While the current statistics are
poor, the presence of these two galaxies in a relatively small survey
area suggests that the abundance of galaxies with such large star
formation rates may have previously been underestimated.
If the high SFR of z8\_GND\_5296 continues down to $z =$ 6.3, it would have a stellar
mass of $\sim$5 $\times$ 10$^{10}$ M\sol, comparable to
the extreme star forming $z =$ 6.34 galaxy HFLS3 (Table 1).\cite{riechers12}
Should z8\_GND\_5296 in fact be a progenitor of such SMGs, it is likely
in the process of enshrouding itself in dust.

Both z8\_GND\_5296 and GN-108036 also have young inferred ages
and IRAC colors indicative of strong [O\,{\sc iii}] emission.
Given the difficulty of detecting Ly$\alpha$ emission at $z \geq$ 6.5, 
it is interesting that these highest redshift Ly$\alpha$-detected galaxies
appear to have extreme SFRs and high [O\,{\sc iii}] emission.  It may
be that a high SFR and/or a high excitation are necessary conditions
for Ly$\alpha$ escape in the distant universe, perhaps through blowing
holes in the ISM, allowing both Ly$\alpha$ and ionizing photons to
escape.  An outflow in the ISM of 200-300 km s$^{-1}$ could clear a
hole in this galaxy in about 3--5 Myr, or perhaps even sooner if the galaxy is
undergoing a merger, which could preferentially clear some lines of
sight for Ly$\alpha$ to escape.

Finally, we examine the lack of detected Ly$\alpha$ lines in our full
dataset.  If the Ly$\alpha$ EW distribution
continues its observed increase\cite{stark11} from 3 $< z <$ 6 out to $z \sim$ 7--8, we
should have detected Ly$\alpha$ emission from 6 galaxies.  Our
single detection rules out this EW distribution at 2.5$\sigma$
significance.  This confirms previous results at $z \sim$
6.5,\cite{kashikawa06,iye06,pentericci11,ono12} but here we probe $z
>$ 7.  The lack of detectable Ly$\alpha$ emission is unlikely to be
due to sample contamination, as contamination by lower redshift interlopers is likely not dominant at $z =$ 7 given the low
contamination rate at $z =$ 6.\cite{pentericci11}  
To explain the low detection rate of Ly$\alpha$, an IGM neutral fraction at
$z =$ 6.5 as high as 60--90\% has been proposed,\cite{pentericci11,ono12}
 implying a rapid increase from $z =$ 6.\cite{fan06}
However, most other observations are consistent with an IGM neutral fraction $\leq$
10\% at $z =$ 7,\cite{bolton11,finkelstein12b} thus alternative
explanations for the dearth of Ly$\alpha$ emission need to be
explored.  

One alternative explanation for at least part of the Ly$\alpha$ deficit may be gas within galaxies. 
A high gas-to-stellar mass ratio may be consistent with the very high SFR of z8\_GND\_5296, as galaxies should not have SFRs (for long periods) exceeding
their average gas accretion rate from the IGM (which is set by the
total baryonic mass).  For the inferred stellar mass and redshift,
z8\_GND\_5296 must have a gas reservoir of about 50 times the stellar
mass to give an accretion rate comparable to the SFR.\cite{dekel13}
If true, this galaxy would have a gas surface density
similar to the most gas-rich galaxies in the local universe, and its
SFR would be consistent with local relations between the gas and SFR surface
densities.\cite{kennicutt12}
The large gas-to-stellar mass ratio could be due to low metallicities at earlier
times which may initially inhibit star-formation allowing the formation of such a large
gas reservoir.\cite{krumholz12}  If such high
gas-to-stellar mass ratios are common amongst $z >$ 7 galaxies, it
could explain the relative paucity of Ly$\alpha$ emission in our
observations.  Direct observations of the gas properties of
distant galaxies are required to make progress understanding both the
fueling of star formation, and the escape of Ly$\alpha$ photons.

\begin{table*}
\begin{center}{\bf Table 1: Ly$\alpha$ Spectroscopically Confirmed Galaxies at $z >$ 7}\end{center}
\begin{center}
\begin{tabular}{cccccc}
\hline
\hline
ID & z$_{Ly\alpha}$ & M$_{UV}$$^{\dagger}$ & Rest EW$_{Ly\alpha}$
(\AA) & SFR (M\sol/yr$^{-1}$) $^{\ddagger}$ & log Stellar Mass (M\sol)\\
\hline
z8\_GND\_5296&7.508&$-$21.2\phantom{$^{a}$}&\phantom{0}8&330\phantom{$^{a}$}&9.0\\
SXDF-NB1006-2\cite{shibuya13}&7.215&$-$22.4$^{a}$&15&\phantom{0}56$^{a}$&---\\
GN-108036\cite{ono12}&7.213&$-$21.8\phantom{$^{a}$}&33&100\phantom{$^{a}$}&8.8\\
BDF-3299\cite{vanzella11}&7.109&$-$20.6\phantom{$^{a}$}&50&\phantom{00}9\phantom{$^{a}$}&---\\ 
A1703\_zD6\cite{schenker12}&7.045&$-$19.4\phantom{$^{a}$}&65&\phantom{00}4\phantom{$^{a}$}&---\\
BDF-521\cite{vanzella11}&7.008&$-$20.6\phantom{$^{a}$}&64&\phantom{00}9\phantom{$^{a}$}&---\\
\hline
IOK-1\cite{iye06,ono12}&6.965&$-$21.6\phantom{$^{a}$}&43&\phantom{0}10\phantom{$^{a}$}&---\\
HFLS3\cite{riechers13}&6.337&---&---&2900\phantom{$^{a}$}&10.6\\
\hline
\hline
\end{tabular}
\end{center}
\caption{Currently known galaxies with
  z$_{Ly\alpha} >$ 7.  We include IOK-1 for comparison, as
  it was the highest-redshift spectroscopically confirmed galaxy for
  several years, and HFLS3, which has the most extreme SFR
  known, and may represent the $z \sim$ 6 evolution of z8\_GND\_5296.  $^{\dagger}$We compute UV absolute
magnitudes for BDF-3299 and BDF-512 using the Ly$\alpha$-corrected
$Y$-band magnitudes, and for A1703\_zD6 using the de-lensed $J$-band magnitude.
$^{\ddagger}$ The SFR for z8\_GND\_5296 and
GN-108036 were both calculated via SED fitting.  The SFR for IOK-1 was
measured from Ly$\alpha$ emission, which is likely a lower limit, due
to unknown absorption.  The SFRs for BDF-3299, BDF-521, A1703\_zD6 and
SXDF-NB1006-2 were calculated from the UV luminosity, which are also likely lower limits,
as the UV luminosity was not corrected for dust attenuation, and the
scaling relation was defined for a stellar population with an age of
100 Myr\cite{kennicutt98}.  The SFR for HFLS3 was derived via the
infrared luminosity.  $^{a}$SXDF-NB1006-2 was only photometrically
detected in a narrowband which encompassed Ly$\alpha$ emission.  The
corresponding UV absolute magnitude, and subsequent SFR, are thus
highly uncertain, with published uncertainties of M$_{UV} = -$22.4 $\!^{+\infty}_{-0.4}$.\cite{shibuya13}}
\end{table*}

\begin{addendum}
\item   [Acknowledgements]  We   thank Mark Dijkstra, James Rhoads and
  Sangeeta Malhotra for useful conversations, as well as Nick Konidaris and Chuck Steidel for
  assistance with the MOSFIRE data reduction pipeline.  We also thank
  our Keck Support Astronomer Greg Wirth for assistance during our
  observing run.  S.L.F.  acknowledges support from  the University of
  Texas at Austin, the McDonald Observatory, and NASA through a NASA
  Keck  PI Data  Award,  administered by  the  NASA Exoplanet  Science
  Institute.  Data  presented herein  were obtained at  the W.  M. Keck
  Observatory   from  telescope   time  allocated   to   the  National
  Aeronautics and Space Administration through the agency’s scientific
  partnership  with the  California  Institute of  Technology and  the
  University of  California. The Observatory was made  possible by the
  generous  financial  support of  the  W.  M.  Keck Foundation.   The
  authors  wish  to recognize  and  acknowledge  the very  significant
  cultural role and reverence that  the summit of Mauna Kea has always
  had within the indigenous  Hawaiian community. We are most fortunate
  to have the opportunity to conduct observations from this mountain.
  This work is also based in part on observations made with the NASA/ESA Hubble Space Telescope,
obtained at the Space Telescope Science
Institute, which is operated by the Association of Universities for
Research in Astronomy, Inc., under NASA contract NAS 5-26555, as well
as the Spitzer Space Telescope, which is operated by the Jet
Propulsion Laboratory, California Institute of Technology under a
contract with NASA.

\item[Author  Contributions]  S.L.F.  wrote  the  text,  obtained  and
  reduced the data, and led  the initial observing proposal.  C.P. and
  M.D. assisted with the analysis of the data.  M.S. and V.T. assisted
  with  the observation planning  and implementation.   K.D.F. performed
  the {\it Spitzer}/IRAC photometry.  A.M.K. was responsible for the reduction
  of the  optical and  near-IR imaging data  used to  select the
  sample.  G.G.F., M.L.N.A. and S.P.W. obtained and reduced  the
  mid-infrared data.  B.J.W. provided grism spectroscopic information.
  B.M., H.C.F., M.G, N.R., A.D., A.F., N.A.G., J.-S. H., D.K., and
  M.R. have contributed in their roles as members
  of   the  CANDELS and S-CANDELS teams,  and  assisted   with  the   planning  and
  interpretation of the observations.

\item[Author  Information]  Reprints  and permissions  information  is
  available  at   www.nature.com/reprints.  The  authors   declare  no
  competing financial interests. Readers are welcome to comment on the
  online  version  of  the  paper.  Correspondence  and  requests  for
  materials should be addressed to S.L.F. (stevenf@astro.as.utexas.edu).

\end{addendum}

\clearpage

\input{SI.tex}

\end{document}

%% file: SI.tex
\setcounter{page}{1}
\setcounter{figure}{0}
\setcounter{table}{0}
\renewcommand{\thefigure}{S\arabic{figure}}
\renewcommand{\thetable}{S\arabic{table}}

\begin{center}
{\bf \Large \uppercase{Supplementary Information} }
\end{center}

\section{Sample Selection}

Our sample of high-redshift galaxies in the GOODS-North field was selected using an updated
version of the criteria presented in previous
papers;$^{15,22}$ the full sample will be
published in Finkelstein et al.\ in prep.  These papers can be
consulted for more details, but here we briefly recap our
process.  

The optical imaging comes  from the  GOODS survey,\cite{giavalisco04}
and we used the v2.0 ACS  imaging, consisting of mosaics in the F435W,
F606W, F775W  and F850LP filters.   The near-infrared data  comes from
the CANDELS survey, and we used the CANDELS
team's early data products (v0.1) in  the F105W, F125W and  F160W filters.
The  CANDELS  survey obtained  data  at  two depths,  denoted  as
``WIDE''  and  ``DEEP''.   The imaging
used here consists of the full depth in the Northeast WIDE region, and
about half  of the  full  depth  of the  DEEP  region.  The  5$\sigma$
limiting  magnitudes, measured  in 0.4\arcsec-diameter  apertures, for
the ACS bands  are: 28.1, 28.3, 27.8 and  27.7 mag, respectively (all
magnitudes are quoted in the AB system\cite{oke83}).  For the
three WFC3 bands,  the existing DEEP 5$\sigma$ depths are  27.9, 27.9
and 27.7 mag,
while  for  the WIDE  region,  the depths  are  27.4,  27.4 and  27.3 mag,
respectively.   Additionally, we add to our analysis new, extremely
deep, optical data obtained with ACS in parallel to the CANDELS
observations.  These data were obtained in the F814W filter, and have
an exposure time of 57,000 s at the position of z8\_GND\_5296, showing no
detectable flux within a 0.4\arcsec-diameter  aperture 5$\sigma$ depth of
28.8.  We   created  photometry  catalogs   with  the  Source
Extractor software,\cite{bertin96}  using a weighted sum  of the F125W
and F160W images as the  detection image.  We measured colors in small
elliptical  apertures,   setting  the  Kron   aperture  parameters  to
Kron$\_$fact$=$1.2  and  min$\_$radius$=$1.7.   Aperture corrections  were
measured  in  the F160W  band  by comparing  the  flux  in this  small
aperture  to  that  in   the  default  MAG$\_$AUTO  aperture,  which  is
representative  of the total  flux.  Photometry  was performed  on the
DEEP and WIDE regions separately.  Photometry errors were obtained by
providing Source Extractor with accurate RMS images.  

No RMS map was available for the F814W data, but we followed the same procedures used 
to calibrate noise maps for the standard CANDELS HST data products\cite{Guo13}.
We measured the RMS and autocorrelation function of the background noise near the position 
of our object, after masking out sources.   We scaled the correlation-corrected RMS to the number
of pixels in the elliptical photometry aperture, finding a total 1$\sigma$ F814W flux uncertainty of 
5 nJy.  This is a factor of about 3$\times$ deeper than the GOODS F775W or F850LP imaging 
at this position.  These F814W data were not available at the time of our observations, so here we
add the F814W non-detection to the photometric redshift and spectral
energy distribution analysis for z8\_GND\_5296, assuming a flux error of 5 nJy.

To select our galaxy sample, we utilized a photometric redshift
fitting technique, using the EAZY software
package\cite{brammer08} to estimate the likely redshift (and
associated redshift probability distribution function, $\mathcal{P}$[z]) by finding the
best-fitting combination of redshifted galaxy spectral templates.
Both our DEEP and WIDE catalogs were run through EAZY.  We then
selected samples with $\Delta z \sim$ 1 centered at $z_{sample} =$ 6, 7 and 8.
Rather than using the best-fit photometric redshift to select our
galaxy sample, we utilized the full redshift probability distribution
function.  For a given object to be in our sample, it had to meet all
of the following criteria:
\begin{itemize}

\item Signal-to-noise in both the F125W and F160W bands $\geq$ 3.5.

\item $\geq$70\% of the integral of $\mathcal{P}$(z) in the primary redshift solution.

\item $\int \mathcal{P}$($z_{sample} \pm 0.5)dz \geq$ 0.25

\item $\int \mathcal{P}$($z_{sample} \pm 0.5) dz \geq$ $\int
  \mathcal{P}$($z_{sample \pm 1} \pm 0.5) dz$

\item $\int \mathcal{P}$($z > [z_{sample}-2]) dz \geq$ 0.5

\item $z_{best}$ $> z_{sample}-2$

\item $\chi^2 < 60$
\end{itemize}
These are very similar to the criteria used in our
previous publications, and they have been shown to produce samples
which match up very well with available spectroscopic redshifts at $z
<$ 7.$^{15}$

The selected sources were visually inspected to reject artifacts
such as diffraction spikes and oversplit regions of bright galaxies.
Additionally, the colors of galaxy candidates were compared to the
expected colors of M, L and T-dwarf stars, and any sources with
star-like colors which were also unresolved were rejected from the
sample.  Finally, the optical bands were also inspected to ensure that
they visually appeared to contain no significant ($>$1.5$\sigma$) flux (in practice,
sources with significant optical flux would have already been rejected
by our selection criteria).  Our final galaxy samples consist of 175
candidate galaxies at $z \approx$ 6, 85 at $z \approx$ 7 and 25 at $z
\approx$ 8.

\section{Spectroscopic Followup Sample}
From our parent sample of candidate galaxies, we selected those for
spectroscopic followup with MOSFIRE via two
criteria: 1) apparent F160W magnitude, and 2) maximizing $\int \mathcal{P}(7.0 < z <
8.2) dz$ (which corresponds to the redshift range placing Ly$\alpha$ in
the MOSFIRE $Y$-band grating).  We first prioritized based on
brightness, and then within each magnitude bin, we prioritized based on
the highest value of the integral defined above.  We input these
catalogs into the MAGMA
software$^\mathrm{F1}$\let\thefootnote\relax\footnote[1]{$^\mathrm{F1}$http://www2.keck.hawaii.edu/inst/mosfire/magma.html}, 
which was created by the MOSFIRE team to design mask configurations.
The software searches a large (user-defined) parameter space in both
right ascension, declination and position angle to maximize the total
priority of sources.  We
designed two masks: GOODSN\_Mask1, with a position angle of 34
degrees, containing 24 candidate high-redshift
galaxies, and GOODSN\_Mask2, with a position angle of $-$9.5
degrees, containing 19.

\section{Observations and Data Reduction}

Our observations took place on UT 18-19 April 2013 under clear, mostly
photometric conditions.  We used MOSFIRE with the $Y$-band grating,
which observes $\sim$0.97 -- 1.12 $\mu$m, and set the slit widths to 0.7\arcsec.
We observed each configuration for one
night, taking 180 sec exposures with an ABAB dither pattern, with
dither positions separated by 2.5\arcsec, yielding
a total exposure time of 5.6 hr for the first configuration and 4.45 hr for the second.
The data were reduced using the MOSFIRE data reduction pipeline$^\mathrm{F2}$\footnote[2]{$^\mathrm{F2}$http://code.google.com/p/mosfire/},
which in brief calculates a wavelength solution using the night sky lines, performs sky
subtraction, flat-fielding and rectification, and saves each two-dimensional
slit spectrum as a single image.  We examined the expected slit
position for each object by eye to search for detected emission lines.
Given our dither pattern, true features are identifiable with a
positive signal and two negative signals on each side in the spatial dimension, due to the sky
subtraction (i.e., each negative signal contains half of the amplitude
of the positive signal).  We identified four plausible emission lines
from the first mask, and four from the second mask.

\subsection{One-Dimensional Spectral Extraction}
For these seven sources, we performed 
one-dimensional spectral extraction with a 1.6\arcsec\ box in the spatial dimension
($\sim$2$\times$ the seeing during the run, which varied from 0.6 --
0.8\arcsec).  
The error spectrum was similarly extracted from the
inverse variance spectrum created by the pipeline.  To ensure that the
error spectrum accurately matched the errors in the object spectrum,
we scaled the error spectrum to be representative of noise variations in
the object spectrum, by measuring the standard deviation in the
signal-to-noise of pixels in regions clear of sky
emission lines, and scaled the error spectrum so that this equaled
unity.  To determine the significance of the eight extracted lines, we fit a Gaussian
function using the MPFIT IDL package$^\mathrm{F3}$\footnote[3]{$^\mathrm{F3}$http://www.physics.wisc.edu/$\sim$craigm/idl/fitting.html}, finding that only one object had a line detected at
$>$ 5$\sigma$ significance.  This object, called z8\_GND\_5296 in our catalog, is measured to have an
emission line at $\lambda$ = 1.0343 $\mu$m with a significance of
7.8$\sigma$ (Figure 1).  Assuming this line is Ly$\alpha$ places this object at
$z =$ 7.5078 $\pm$ 0.0004, making this the highest redshift galaxy
that has been spectroscopically confirmed via Ly$\alpha$ to date. 
There has been a published spectroscopic confirmation of the gamma ray burst
(GRB) 090423 at $z =$ 8.2, confirmed via continuum spectroscopy of the
Lyman break\cite{tanvir09,salvaterra09}, though due to its very nature this object cannot be
re-observed.  Additionally, although a spectroscopic redshift
of $z=8.56$ for a galaxy has been claimed,\cite{lehnert10}
subsequent observations have shown this to be
spurious.\cite{bunker13}
The properties of z8\_GND\_5296 are summarized in Table S1.  
We note that the uncertainty on the redshift denotes the uncertainty on
centroiding the line.  However, as seen at lower redshift, Ly$\alpha$
is frequently detected at 200--800 km s$^{-1}$ redward of the systemic
redshift\cite{shapley03,steidel10}, thus the systemic redshift for this system may be a few
hundred km s$^{-1}$ lower.

The MOSFIRE data reduction pipeline provides a nominal estimate for the
central row for each objects spectrum, accounting for differing
vertical positions in the slit.  To ensure that our extracted emission
line in the spectrum of z8\_GND\_5296 is in the correct spatial position, we used three
sources in our mask with well-detected continuum; one was a star,
while the other two were $z \sim$ 1 galaxies placed in the mask as
fillers.  We found that all three sources had centroids $\sim$4--5
pixels below the pipeline estimate, with a mean offset of 4.7 pixels.
Examining the emission line in the 2D spectrum of z8\_GND\_5296, we
find that this line also has a centroid offset from the pipeline estimate by
4.7 pixels.  Thus, we conclude that the observed emission line is at
the expected position for the high-redshift galaxy we intended to
observe, and we use this offset position as the extraction center.  
As shown in Figure 2, there are no other sources in the
slit, though there are two galaxies located 2.3\arcsec\ and
3.2\arcsec\ southwest of our target.  The closer galaxy would lie
1.1\arcsec\ off the slit center, and would be offset by 1.9\arcsec\ along
the slit from our object, which corresponds to $\sim$10 pixels in the
2D spectrum.  Any emission from these objects which happened to fall
in the slit would thus be clearly separated from our observed emission
line.  In Section S4.3, we find that both of these nearby galaxies have
spectroscopic redshifts of 0.39, which would not place any known
emission line near 1.0343 $\mu$m.

In order to examine the possibility of a false positive detection, we
examined the signal-to-noise spectrum, smoothed by the velocity width
of our spectrum, and scaled it such that the value at the peak of our detected line
is equal to the integrated signal-to-noise of the line of 7.8, as
illustrated in Figure S1.  We searched the entire spectrum for apparently significant negative
features; these would be due to noise, and the lack of such features
provides greater confidence that our observation represents a true
emission line from the object z8\_GND\_5296, while the lack of
positive features other than our identified Ly$\alpha$ line provides
further confidence that the line is in fact Ly$\alpha$.

\begin{table*}
\begin{center}{\bf Table S1: Summary of Spectroscopically Confirmed Galaxy}\end{center}
\begin{center}
\begin{tabular}{lccccccc}
\hline
\hline
Object & R.A.\ & Dec & mag$_{F160W}$ & Photo-z 68\% C.L.\ & $\lambda_{line}$ & $z_{Ly\alpha}$ & SNR$_{line}$\\
           & (J2000) & (J2000) & (AB) &     & (\AA) & &   \\
\hline
z8\_GND\_5296&12:36:37.90&62:18:08.5&25.6&7.5 -- 7.9&10342.6&7.508&7.8\\
\hline
\end{tabular}
\end{center}
\vspace{-5mm}
\end{table*}

\subsection{Flux Calibration}
We flux calibrated the spectrum of z8\_GND\_5296 using
observations of the standard star HIP~56157, with a spectral type
of A0V, which we observed in a single long slit directly before our
science observations during our first night of observing.  We obtained
four spectra of this star with an ABAB dither pattern, with each
exposure consisting of 10 2s co-additions, to guard against persistence
and non-linearity.  These observations were reduced in the same manner
as our masks described above, and extracted into a one-dimensional
spectrum with the same size extraction box as that used on our primary
observations.

We derived the flux calibration array by taking a A0V Kurucz\cite{kurucz93}
model spectrum, and scaling it to match the integrated 2MASS magnitude
for this star, interpolating among the $J$, $H$, and $K$ 2MASS
magnitudes to obtain the magnitude appropriate for our spectral range at 1.05
$\mu$m ($m_{1.05 \mu m, AB} =$ 8.08).  We then created a calibration
array by dividing this scaled model spectrum by our observed spectrum,
interpolating over intrinsic stellar absorption features common to
both spectra.  The final array was then multiplied by our object
spectrum (both normalized by their respective exposure times), which
both flux calibrated our object spectrum, and corrected for telluric
absorption features.  Nominally, this procedure also corrects for slit
losses, but only in the case when the seeing during both the standard
and object observations was the same.  In our case, the seeing
was moderately different; the median seeing
during the mask observations was 0.65\arcsec, while it was
0.85\arcsec\ during the standard observations.  Thus an additional
aperture correction of 1.22 was applied to account for the seeing differences.

We measured the line flux of our detected emission line by again
fitting a Gaussian with MPFIT, only now to the calibrated spectrum.
We measured a line flux of 2.64 $\times$ 10$^{-18}$ erg s$^{-1}$
cm$^{-2}$.  We had expected to achieve a 5$\sigma$ limiting line flux
of 2.1  $\times$ 10$^{-18}$ erg s$^{-1}$
cm$^{-2}$ in 5.5 hr (scaled from our initial expectation of 2.0 $\times$ 10$^{-18}$ erg s$^{-1}$
cm$^{-2}$ in 6 hr).  Given our measured line flux, and signal-to-noise
reported above of 7.8, this would imply a 5$\sigma$ limiting line flux
of 1.7 $\times$ 10$^{-18}$ erg s$^{-1}$
cm$^{-2}$.  While this may be the case, there is an additional 
systematic uncertainty in our flux calibration, as the counts varied by $\sim$15\% in the four individual
observations of the standard star.  
Taking this
into account, our measured line flux is 2.64 $\pm$
0.34 (photometric) $\pm$ 0.40 (systematic) $\times 10^{-18}$ erg s$^{-1}$
cm$^{-2}$.  Accounting for the systematic uncertainty, our measured
line flux is consistent with that expected for a signal-to-noise$=$7.8 detection
at $\sim$1.2$\sigma$.  The flux calibration does not have an
impact on our primary science results, but we will use this calibrated
line flux below when discussing the Ly$\alpha$ equivalent width.

 \begin{figure}
\vspace{-1mm}
\centerline{
\includegraphics[width=0.52\textwidth]{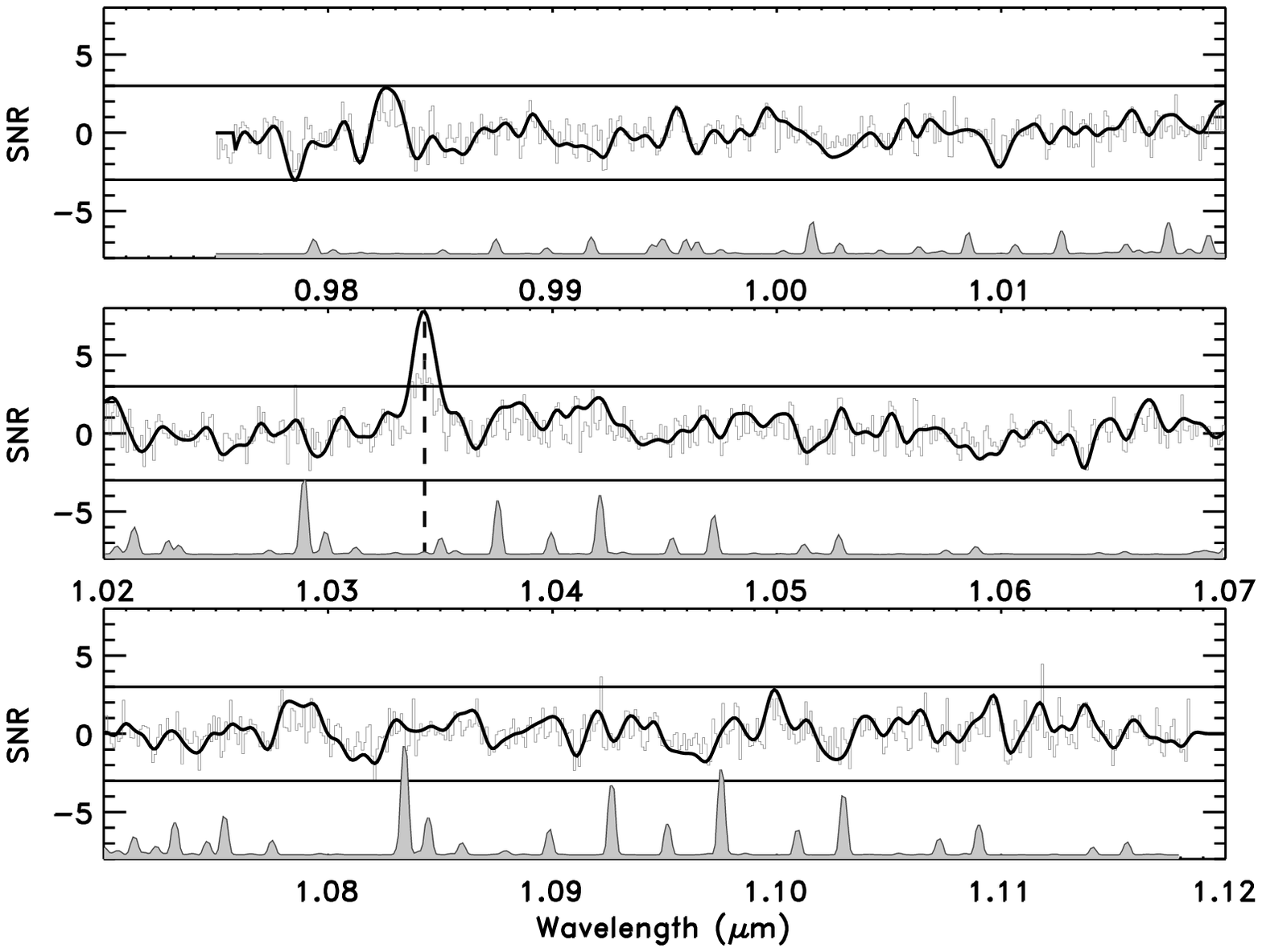}
}
\vspace{-4mm}
\caption{\textbf{Emission line signal-to-noise test}. The results of a signal-to-noise test
    for the one-dimensional spectrum of z8\_GND\_5296 (each row
    represents a different region of the spectrum).  We divided the
  object spectrum by the error spectrum, smoothed by
  the velocity width of our observed line, and normalized the result so
  that the value at the peak of the Ly$\alpha$ line equaled the
  measured integrated line signal-to-noise of 7.8.  The horizontal
  lines denote the $\pm$3$\sigma$ points, and the gray filled spectrum
denotes the (arbitrarily scaled) sky emission.  Only the detected
emission line has a $|$signal-to-noise$|$ $>$3; the absence of
negative fluctuations at this level, which would be due to noise, gives confidence in
the real nature of this emission line.}
\vspace{-5mm}
\end{figure}

\section{Line Identification}

Although our photometric redshift
favors Ly$\alpha$ as the identification for our detected emission line
in the spectrum of z8\_GND\_5296, here we examine the alternatives.
Other plausible alternatives to Ly$\alpha$ (i.e., lines that have been observed to
be reasonably strong at high-redshift) are [O\,{\sc ii}]
$\lambda\lambda$ 3726,3729, H$\beta$ $\lambda$4861, [O\,{\sc iii}]
$\lambda$4959, [O\,{\sc iii}]
$\lambda$5007,  and H$\alpha$ $\lambda$6563 (other lines are possible
if the object is an AGN, but this is
not likely due to the lack of X-ray or long-wavelength detections; see
below).  Of these alternatives,
H$\beta$ and [O\,{\sc iii}] can be ruled out, as if our detected
line was one of these three lines, the remaining two lines should be observed as well.
Specifically, were our observed line [O\,{\sc iii}]
$\lambda$5007, we would expect to see [O\,{\sc iii}] $\lambda$4959 at 10243.5
\AA, which is a region clear of sky emission.  We simulated a
[O\,{\sc iii}] $\lambda$4959 line at this position in our spectrum, with a line
strength 2.98$\times$ less than that of the [O\,{\sc iii}] $\lambda$5007
line,\cite{storey00} and found that such a line would
have been detected at 4.1$\sigma$, thus we rule out [O\,{\sc iii}]
$\lambda$5007 as the identification of our detected line.
Additionally, we can rule
out both [O\,{\sc iii}] lines, as well as H$\alpha$, as they are not located near
strong continuum breaks.  As seen in
Figure 3, we have detected a large photometric break at $\lambda \sim$
1$\mu$m.  We interpret it as the Lyman break, but it could also be the
Balmer break at 3646 \AA\ due to a combination of the high-order Balmer
series transitions, or the 4000 \AA\ break due to metal absorption lines common in older
stellar populations.  Were this the case, then the detected line would
be [O\,{\sc ii}].  

As [OII] is a doublet, we examine the spectrum for signs of the second
line.  The ratio of the $\lambda$3726/$\lambda$3729 line strength
varies from $\sim$0.5--1.5 in H\,{\sc ii} regions, with a typical
ratio of order unity.\cite{osterbrock89}  If our detected line was the red side of the doublet (at 3729
\AA\ rest), we should detect the bluer line at 10334 \AA\ (which is a
clean region) at $>$10$\sigma$, and no line is seen.  If the detected line is the bluer
side of the doublet, then we would expect to see the redder line at
10351 \AA.  This would be directly under the sky line just to the red
of our detected line, which hampers our ability to discern its
presence.  However, given the width of our detected line, if there
was a second line under the sky line, we would expect to see excess
flux just to the red side of the sky residual (i.e., the true line would
be broader than the sky residual), in between the two sky lines.  As
shown in Figure S2, for line ratios of unity or less, the observed
spectrum can rule out the presence of the redder [O\,{\sc ii}] line.
If the $\lambda$3726/$\lambda$3729 ratio is high; close to 1.5, then
it becomes harder to rule out the presence of this line.  However,
there should still be excess flux over what is observed on either side
of the sky line residual -- in particular, on the red side of the sky
line, we would have expected to see emission line flux at the
$\sim$2$\sigma$ level.  Given the lack of detectable flux in this
region, we conclude that the line is unlikely
to be [O\,{\sc ii}].  However, given the unknown strength of any
potential 3729 \AA\ line, in the following we examine further evidence to
differentiate between Ly$\alpha$ and [O\,{\sc ii}].

\begin{figure}
\centerline{
\includegraphics[width=0.48\textwidth]{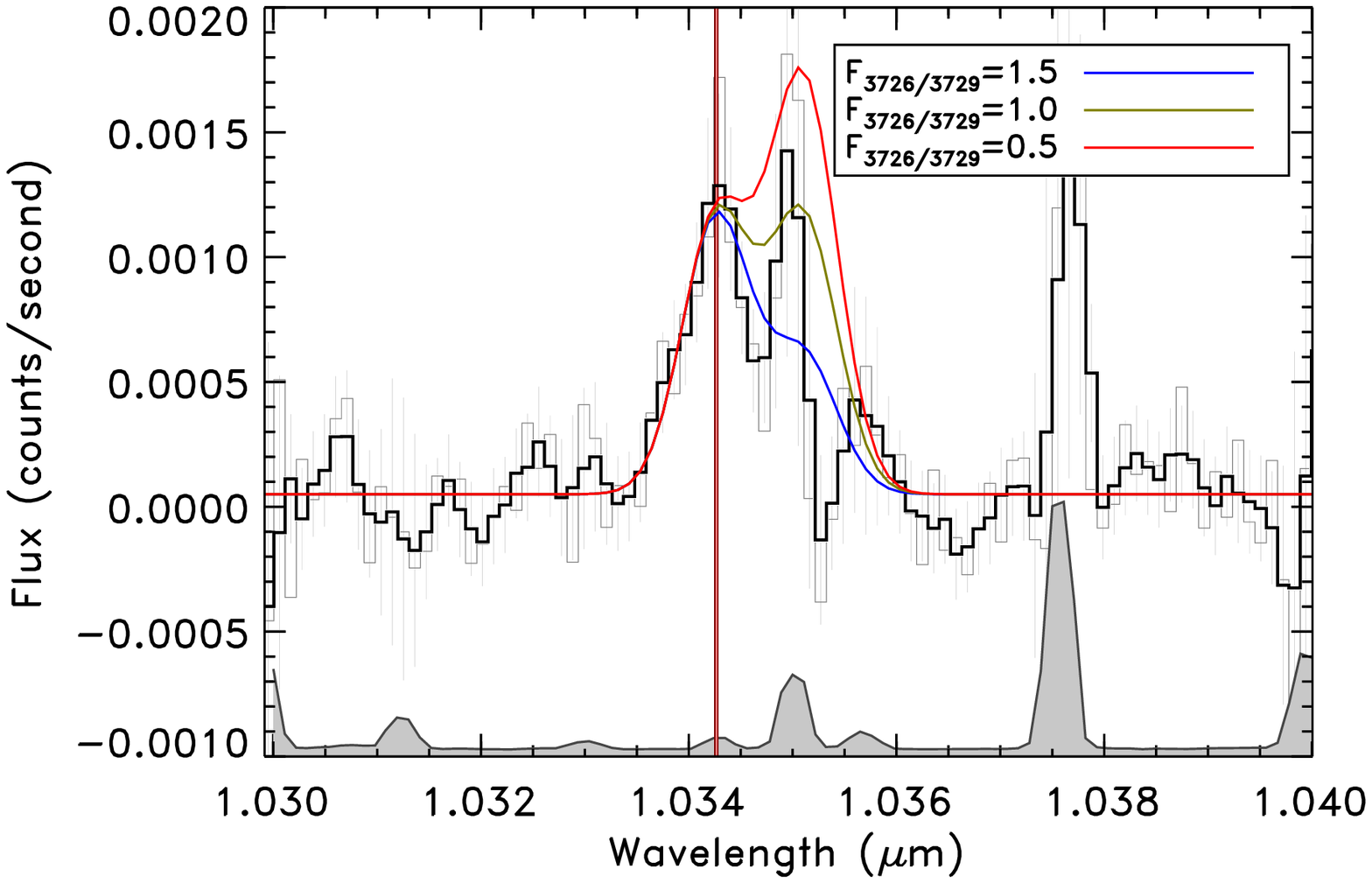}
}
\vspace{-2mm}
\hspace{10mm}
\caption{\textbf{[O\,{\sc ii}] Doublet}. A zoomed in view of our
  source spectrum, overplotting hypothetical [O\,{\sc ii}] doublet
  lines for three values of the ratio between the line fluxes.  Unless
the [O\,{\sc ii}] 3729 \AA\ line is substantially weaker than the 3726
\AA\ line, we would have expected to see highly significant flux from
the redder line.  Even in the case where the redder line is 50\% the
strength of the bluer line, we should still have detected emission
line flux redward of the sky line residual at the $\sim$2$\sigma$ level.}
\vspace{-5mm}
\end{figure}

\subsection{Line Asymmetry}
Another feature which could confirm the Ly$\alpha$ nature of this line
would be any measured asymmetry.  Ly$\alpha$ at high redshift is frequently observed to
be asymmetric,$^{1,3}$ though it has been observed to
be symmetric as well.\cite{rhoads12}  It is assumed that the
asymmetry is caused by absorption of the blue half of the line by
neutral hydrogen in the IGM.  However, a few lines of evidence imply that internal
processes in the galaxy may dominate the observed line profile.
First, Ly$\alpha$ lines at $z \sim$ 2--3 have been observed to be
asymmetric, at an epoch where the IGM absorption is much less.
Second, also at $z \sim$ 2--3, where the systemic redshift can be
measured via rest-frame optical nebular lines, Ly$\alpha$ is seen to
reside $\sim$200-400 km s$^{-1}$ to the red of the systemic
redshift.\cite{steidel10,mclinden11,finkelstein11a,hashimoto13}  This is
likely a result of interstellar winds driven
by intense star-formation, as Ly$\alpha$ photons will preferentially
escape after they have gained some net redshift, and are thus no
longer resonantly scattered.  This enables them to
pass through neutral hydrogen both within the galaxy as well as in
the IGM.  Simulations of galaxies at $z >$ 8 show that
with a wind velocity of $\sim$ a few hundred km s$^{-1}$, not only can
Ly$\alpha$ emission be detectable from a mostly-neutral epoch, but it
can be observed with a symmetric profile.\cite{dijkstra11}  The large
inferred SFR of our object is consistent with this scenario, as it is
very likely driving a strong wind in the interstellar medium.  
Ly$\alpha$ is also symmetric in another bright ($m_{UV}=$25.75) galaxy at
z=6.944,\cite{rhoads12} perhaps indicating that strong star-formation
driven winds are common in these very luminous objects.

The asymmetry of our observed line is
difficult to measure, given the night sky line residual to the
red-side of our line. 
We measure the asymmetry of our emission line by fitting an
asymmetric Gaussian function to the line profile, where the $\sigma$
values on the blue and red side of line center are allowed to be
different.  We then quantify the asymmetry as the ratio of
$\sigma_{red}/\sigma_{blue}$, measuring this ratio to be 1.2 $\pm$
1.4, thus the measured asymmetry is of no significance.
As a further test of our ability to measure any asymmetry in the
detected emission line, we ran a series of simulations, placing mock
emission lines with the same integrated line flux as our measured
line, but with a known value of asymmetry, in our one-dimensional spectra.  We investigated
asymmetry values of both 2.0 and 1.5, and we placed these mock lines at
three locations: 11082.6, 10119.4 and 10250.0 \AA.  The first two
locations correspond to regions 7.4 \AA\ blueward of a skyline with a
similar amplitude to the skyline 7.4\AA\ redward of our detected
emission line; the first of these two has a positive sky-subtraction
residual, while the second has a negative residual.  The third
wavelength is a region with no sky emission lines.  In each of these
six simulations, the measured asymmetry was consistent with unity
(i.e., a symmetric line) at $\sim$1$\sigma$.  The measured asymmetry
values and associated uncertainties were 3.2 $\pm$ 2.8, 3.2 $\pm$ 2.1
and 3.9 $\pm$ 2.6 for the simulations where the input asymmetry value
was 2.0, and 2.8 $\pm$ 4.7, 3.2 $\pm$ 2.0 and 2.3 $\pm$ 2.1 where the
input asymmetry value was 1.5.  Although each of these simulations
results in a mean asymmetry value greater than unity, the very large
uncertainties imply that our spectra are not of high enough
signal-to-noise to detect a moderate amount of asymmetry were it
present in the detected emission line.  Deeper spectra with higher
spectral resolution may make this possible, but given the presence of
sky emission lines around our detected object, it may yet prove
difficult.  We thus conclude that we cannot rule out moderate
asymmetry in our detected emission line.

\begin{figure}
\centerline{
\includegraphics[width=0.48\textwidth]{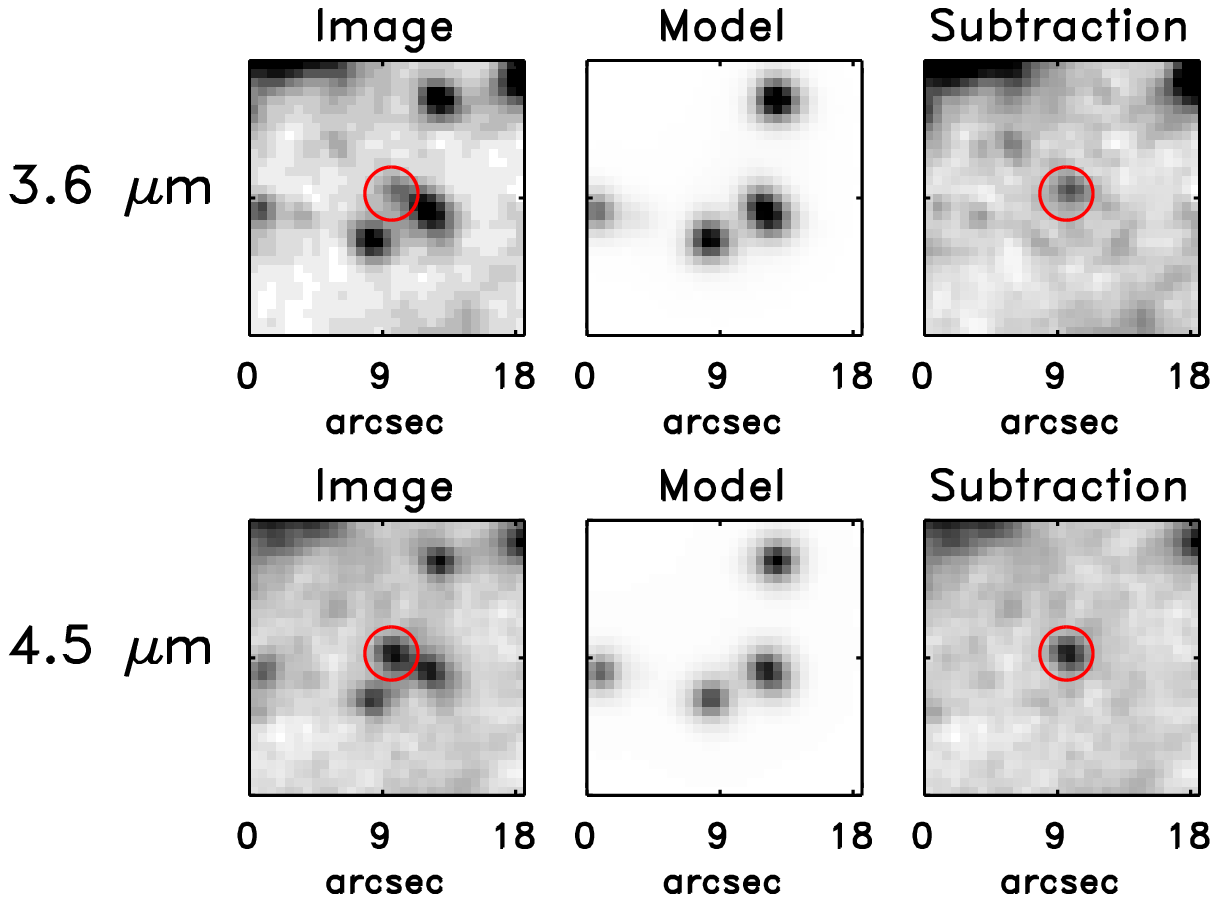}
}
\vspace{-4mm}
\caption{\textbf{IRAC photometry}. 18.6\arcsec\ stamps of
  z8\_GND\_5296 in the IRAC 3.6 (top row) and 4.5 (bottom row) $\mu$m
  bands, highlighting the de-blending algorithm we used to perform our
  source photometry.  The first column is the image, the second is the
  GALFIT source model of nearby sources, and the third is the model-subtracted image,
  which clearly shows a significant detection for z8\_GND\_5296 in
  both bands, with minimal residuals from other sources.  This fitting
  was straightforward, as the neighbors are relatively
  faint, and are well fit by point-sources.  When we
  performed photometry, z8\_GND\_5296 was included in the GALFIT model, thus
  the quoted magnitudes come from this point-source fitting method
  rather than a less accurate circular aperture.}
\vspace{-5mm}
\end{figure}

\subsection{Equivalent Width}
Photometric surveys for Ly$\alpha$ emitting galaxies at high redshift using
narrowband filters frequently use the equivalent width of the line as
a method to remove [O\,{\sc ii}] emitting
``contaminants''.\cite{rhoads00,gawiser06b}  The dividing line used is
commonly 20 \AA\ in the rest frame of Ly$\alpha$. 
As we do not detect the continuum in our spectrum, we must use the
photometry to derive the continuum level near the detected emission
line.  We use the best-fitting model from our SED fitting (see the
next subsection), to derive the continuum flux density just redward of
Ly$\alpha$ (at a
rest-frame wavelength of 1225 \AA), which we find to be 4.15
$\times$ 10$^{-20}$ erg s$^{-1}$ cm$^{-2}$ \AA$^{-1}$.  The EW is then
defined as the ratio of the line flux to the continuum level, which we
find to be 64 $\pm$ 8 (photometric) $\pm$ 10 (systematic) \AA.  If the
line is Ly$\alpha$ at $z =$ 7.51, this would correspond to a
rest-frame EW $=$ 7.5 \AA, while for [O\,{\sc ii}] at $z =$ 1.78, the
rest-frame EW would be 23 \AA.  An emission line of this small EW would have
a negligible impact on the integrated F105W magnitude, and it does not provide
further evidence excluding the possibility of [O\,{\sc ii}], although it
does support our primary conclusion that the equivalent width
distribution at $z >$ 7 has been drastically reduced.

\setcounter{table}{1}
\begin{table*}
\begin{center}{\bf Table S2: Measured Broadband Flux Densities of z8\_GND\_5296}\end{center}
\begin{center}
\begin{tabular}{cccccccccc}
\hline
\hline
F435W&F606W&F775W&F814W&F850LP&F105W&F125W&F160W&3.6$\mu$m&4.5$\mu$m\\
 \hline
$-$5.4 $\pm$ 10.5&$-$5.0 $\pm$ 8.5&17.8 $\pm$ 13.9&0.0 $\pm$ 5.04&0.7 $\pm$ 16.0&102 $\pm$ 10&194 $\pm$ 12&218 $\pm$ 14&256 $\pm$ 25&631 $\pm$ 51\\
\hline
\end{tabular}
\end{center}
\vspace{-5mm}
\caption{All fluxes are in nJy (10$^{-32}$ erg s$^{-1}$ cm$^{-2}$
  Hz$^{-1}$).  While the measured signal-to-noise in the F775W band is
  1.3, the lack of detections in all other optical bands (including the
  stacked optical image) as well as in a smaller circular aperture implies that this is due to random noise.}
\end{table*}

\subsection{Grism Spectroscopy and Lensing}
Most of the GOODS-North field,  including the region of interest here,
has been  observed with {\it HST} WFC3 infrared  slitless grism spectroscopy
(Weiner  et al.\  in preparation),  covering the  1.1--1.65$\,\mu$m spectral
range.  This range does not include the line at 1.0343$\,\mu$m that we
observe with  MOSFIRE, but if that line was [O\,{\sc ii}] at $z  = 1.78$ or
[O\,{\sc iii}] at  $z = 1.07$, other emission  lines (namely, [O\,{\sc iii}]+H$\beta$
or H$\alpha$+[N\,{\sc ii}], respectively) would fall within the grism spectral
range.  These are not observed, to an approximate 3$\sigma$ flux limit
of $3 \times 10^{-17}$~erg~s$^{-1}$~cm$^{-2}$, neither in the spectrum
of the faint galaxy z8\_GND\_5296, nor in the two galaxies that fall a
few arcseconds  away to the southwest,  near (but not  on) the MOSFIRE
slit (see Section S3).  The  closer (northeastern) of  these two galaxies has  a secure
Keck  DEIMOS spectroscopic  redshift $z  =  0.387$ (Stern  et al.\  in
preparation),  which would  not  place any  strong  emission lines  at
1.0343$\,\mu$m.  The second  (southwestern) galaxy has no ground-based
spectroscopy  to our  knowledge.  Spectral  templates cross-correlated
with the WFC3 grism spectrum of this southwestern galaxy yield a possible
redshift $z  = 0.39 \pm  0.01$, largely due  to a feature  that would
correspond  to  the  [S\,{\sc iii}]~$\lambda$9069\AA\  emission line  at  that
redshift.   While quite tentative,  this is  also consistent  with the
secure and accurate Keck redshift  for the northeastern galaxy that is
about  1  arcsec away,  suggesting  that the  two  may  be a  physical
pair.  In any case, there is  no evidence to favor (and several reasons
to discount) the possibility that  the MOSFIRE emission line is due to
contamination from a nearby foreground galaxy.

This nearby pair of galaxies is unlikely to act as a significant
gravitational lens.  At the spectroscopic redshift of $z =$ 0.39 for both galaxies, we measure
stellar masses from SED fitting for the NE galaxy of 5.8 $\times$
10$^{7}$ M\sol, and for the SW galaxy of 1.7 $\times$ 10$^{7}$ M\sol.  To determine whether
these could plausibly magnify our $z =$ 7.51 galaxy, we compute their
Einstein radius, assuming a lens redshift of $z =$ 0.39, and a source
redshift of $z =$ 7.51.  For this calculation, we require the total mass of
the galaxies, including dark matter, which we conservatively assume is
10$\times$ the stellar mass (cf.\ compare to samples of massive
galaxies in strong lensing surveys that find stellar-mass fractions of
50-100\% within the Einstein radius\cite{koopmans06}).  For the NE
galaxy, we find an Einstein
radius of 0.05\arcsec, while for the SE galaxy we find an Einstein
radius of 0.03\arcsec.  The separation between these sources and
z8\_GND\_5296 is $\sim$2.3\arcsec\ and 3.2\arcsec, respectively.
Additionally, even the largest Einstein radius from the strong-lensing galaxies of Sloan Lens ACS
Survey\cite{koopmans06} would reach only 1.3\arcsec\ at $z =$ 0.39 (for a lensing galaxy with 
stellar mass $>$10$^{11}$ M\sol; more than 100 times that of the
$z=$0.39 galaxies here).  We thus conclude
that strong gravitational lensing is not affecting the inferred luminosity.

\subsection{Spectral Energy Distribution Fitting}

In the above subsections, we have attempted to discern between the
[O\,{\sc ii}] and Ly$\alpha$ identification of the detected emission line by
looking at the line properties itself.  However, the strongest
evidence either way can likely be had by looking at the full
photometric SED.  Although an emission line near a spectral break can be indicative of
both Ly$\alpha$ or [O\,{\sc ii}], the stellar populations which would
create these signatures would be drastically different.  
We utilized the same {\it HST} photometry that went into the
photometric redshift fitting, only now we also added in {\it
  Spitzer}/IRAC data at 3.6 and 4.5$\mu$m.  
We utilized new IRAC data from the {\sl Spitzer} Very Deep Survey of the HST/CANDELS
fields (S-CANDELS; PI Fazio), which is a Cycle 8 {\sl Spitzer}/IRAC program to cover the CANDELS wide fields (0.2\,deg$^2$) with a total integration
time of $\sim$50 hr in both IRAC bands at 3.6 and 4.5\,$\mu$m.  S-CANDELS
data acquisition in the CANDELS GOODS-N field was completed over the
course of two visits, during 2012 January and 2012 July.  The  data
were reduced to mosaic form following procedures identical to those
described for the coextensive, wider but shallower {\sl Spitzer}
Extended Deep Survey.\cite{ashby13}  At the position of z8\_GND\_5296,
the exact integration times are 47.2 and 57.8 hr in the 3.6 and 4.5
$\mu$m bands, respectively.  A rms image was produced for each band by taking the
inverse of the square root of the coverage map, and scaling it so that
the mean value was equal to the mean of the pixel-to-pixel
fluctuations in empty regions of the image.

As shown in Figure S3, z8\_GND\_5296 is clearly detected in both bands,
but due to the large beam of {\it Spitzer}/IRAC, simple aperture
photometry will result in inaccurate fluxes due to contamination from
nearby neighbors.  
We therefore fit and subtracted nearby sources in a $19\arcsec \times
19\arcsec$ region around z8\_GND\_5296 in each of the IRAC images.
Positions, magnitudes, and radial profiles of the sources in this
region were derived by running Source Extractor on the higher
resolution {\it HST} F160W-band images.  Each source found, including z8\_GND\_5296, was modeled
on the IRAC images with the galaxy-fitting software package
GALFIT\cite{peng10} (v3.0) in a manner similar to our previous
work.\cite{kfinkelstein11}  Figure~S3 illustrates the process.  GALFIT
requires a point-spread function
(PSF), which was constructed using stars in the large IRAC
mosaics. The FWHMs of the IRAC PSFs were 1.9\arcsec.  The extracted AB
magnitudes of z8\_GND\_5296 are $m_{3.6} = 25.38 \pm 0.09$ and
$m_{4.5} = 24.40 \pm 0.07$.  We note that these photometric errors
include the uncertainty due to deblending, which we verified by
varying the neighbor fluxes within their 1$\sigma$ uncertainties, and
noted that it changed the flux of the galaxy of interest by $\leq$ 9\%
for the 3.6 $\mu$m band, and $\leq$5\%
for the 4.5 $\mu$m band; both at or less than the quoted photometric uncertainties.
As expected from inspecting the image, the 4.5~$\mu$m flux is much
brighter, which we will comment on below.
We also included constraints during SED fitting at 5.8 and
8.0 $\mu$m, using images from the GOODS {\it Spitzer} survey.  There was no
significant flux at the position of z8\_GND\_5296 (as expected for a
source at high redshift), thus during the SED fitting, these fluxes
were set to zero, and the flux errors were set to the 1$\sigma$ limit
of the images, which are AB magnitudes of 23.485 for 5.8 $\mu$m and
23.355 for 8.0 $\mu$m.  These
limits are 1-2 mag brighter than both of our best-fit models.  The
photometry of z8\_GND\_5292 is listed in Table S2.

We compared the 12 photometric points of our SED to a suite of
stellar population models, using the updated models of Bruzual \&
Charlot.\cite{bruzual03}  In these models, we assumed a Salpeter
initial mass function, and varied the stellar
population age, metallicity, dust content and star-formation history.  
There is mounting evidence that a dust attenuation law,
$A(\lambda)/E(B-V)$, similar to that derived for the Small Magellanic
Cloud (SMC) better reproduces the UV-optical colors and IR/UV ratios
for young, presumably lower-metallicity galaxies at high redshifts,\cite{reddy12,tilvi13,oesch13b}
 compared to the dust attenuation
law for local UV-luminous starbursts\cite{calzetti00} that is more commonly
used (see discussion in Tilvi et al.\cite{tilvi13}).    This is perhaps unsurprising as
the SMC is frequently pointed to as a local analog for high-redshift
galaxies.  We thus use the SMC dust-attenuation curve derived by Pei\cite{pei92} to model the effects of dust on our model spectra.
Additionally, recent evidence implies that high-redshift galaxies likely have a
rising star-formation history on
average,\cite{papovich11,finlator11,reddy12,jaacks12} thus we allow both
exponentially rising and declining star-formation histories.  The
stellar mass is found as the normalization between the observed fluxes
and the best-fit model.  We include nebular emission lines using the
emission line ratios published by Inoue et al. (see also Salmon et al., in prep).\cite{inoue11}  The
best-fitting model is found via $\chi^2$ minimization, and the
uncertainties on the best-fitting parameters are found via Monte Carlo
simulations, varying the observed fluxes by an amount proportional to
their photometric errors.  This
procedure is similar to that used in our previous work, to which we
refer the reader for more details.$^{15,22}$

We perform two fits; first fixing the redshift to $z =$ 7.51 should
our detected line be Ly$\alpha$, and secondly fixing $z =$ 1.78, if
the line were [O\,{\sc ii}].  We note that in the high-redshift fit,
we exclude the $Y$-band photometry, as the highly-star-forming nature
of this object implies that it likely has strong intrinsic Ly$\alpha$
emission, which will be included in the models.  However, given the
weak Ly$\alpha$ flux observed, the emission is likely being attenuated
by gas somewhere along the line-of-sight; this effect is not included
in the modeling.
As briefly discussed in the main text,
the observed photometry of this source is much more consistent with a
redshift of 7.51, and thus a line identification of Ly$\alpha$
(reduced $\chi^2_r[z=7.51] =$ 0.8 and $\chi^2_r[z=1.78] =$ 14.7).
This is primarily due to two wavelength regimes, highlighted by the
right panel of Figure 3, which shows the values of $\chi^2$ for each
band and redshift.  First, z8\_GND\_5296
is completely undetected in the
optical, even in the ultradeep F814W band.  As can be seen in Figure 3, the lack of a significant detection in
the optical strongly favors the high-redshift solution, with
$\Delta\chi^2$ ($\chi^2_{z=1.78} - \chi^2_{z=7.51}$) $=$ 2.0, 17.6 and 3.4
for the F606W, F814W and F850LP bands, respectively (the F775W band is less
discerning, as it has a formal 1.3$\sigma$ detection; due to the
non-detections in the surrounding bands and in the stack of
all optical bands, as well as the non-detection in this band in a
smaller circular aperture, we attribute this to random noise).  Second, the IRAC bands also strongly favor
the high-redshift solution, with $\Delta\chi^2 =$ 19.1 and 17.4 for the
3.6 and 4.5$\mu$m bands, respectively.  This is understandable as at
$z =$ 7.5, [O\,{\sc iii}] is located in the 4.5 $\mu$m band, and a
strong emission line could create the observed color.  At $z =$ 1.78,
there is no such strong emission line in this band, thus the models
struggle to fit the observed color.  As we discussed in the main text,
the inferred [O\,{\sc iii}] EW can be used to diagnose the metallicity
of this galaxy.  We quote the [O\,{\sc iii}] EW as that from the
best-fitting model, with the quoted 68\% and 95\% confidence ranges
coming from the Monte Carlo simulations (Table S3).

\begin{table*}
\begin{center}{\bf Table S3: 68\% Confidence Range of Physical Properties for z8\_GND\_5296}\end{center}
\begin{center}
\begin{tabular}{cccccc}
\hline
\hline
z & Stellar Mass & Age & E(B-V)  & SFR (t $<$ 10 Myr) & EW ([O\,{\sc iii}])\\
           & (M\sol) & (Myr) &   & (M\sol\ yr$^{-1}$) & (\AA) \\
\hline
7.51&0.9 -- 1.2 $\times$ 10$^{9}$&1 -- 3&0.12 -- 0.18&320 -- 1040&560 -- 640\\
1.78&1.6 -- 1.8 $\times$ 10$^{9}$&510 -- 570&0.0 -- 0.0&0 -- 0&---\\
\hline
\end{tabular}
\end{center}
\caption{The values given correspond to the 68\% confidence range for
  the quoted parameters.  The initial mass function (IMF) was assumed
  to be Salpeter; were it of a Chabrier form, the stellar masses and
  star-formation rates would be lower by a factor of 1.8.}
\end{table*}

The SFR quoted in Table S3 is a time averaged SFR.  For models where
the stellar population age is older than 100 Myr, we integrate the
star-formation history over the past 100 Myr to determine the SFR.
For younger populations, we simply divide the stellar mass by the
stellar population age.
This time-averaged SFR is extremely high for our $z =$ 7.51 fit, with a
68\% confidence range from 320 -- 1040 M\sol\ yr$^{-1}$ (best-fit $=$
330 M\sol\ yr$^{-1}$).  Given the
observed photometry, this is plausible, as the bright rest-frame UV
coupled with strong inferred [O\,{\sc iii}] emission drives the fit to
a young age (3 Myr in this case), when the SFR will be very high.  
However, any SED-fitting-based SFR for very young ages will be extremely
sensitive to very short timescale variations in the SFR that are
extremely difficult to constrain, thus the inferred SFR has a large uncertainty.
Although a young age is necessary to reproduce the inferred [O\,{\sc
  iii}] EW, metallicity will also have a strong effect on the [O\,{\sc
  iii}] EW, and we only coarsely sample the metallicity.  To see what
constraints we can place on the SFR without requiring assumptions on
the [O\,{\sc iii}] line, we performed another fit to the data,
excluding the IRAC 4.5 $\mu$m band.  In this fit, the time-averaged
SFR ranges from 120 -- 530 M\sol\ yr$^{-1}$ (best-fit $=$
260 M\sol\ yr$^{-1}$).  Thus, even without allowing the [O\,{\sc iii}] emission to influence
our fit, this galaxy still has an extremely high time-averaged SFR.

As one final check, we calculate the SFR using the UV-luminosity to
SFR conversion published by Kennicutt et al.,$^{27}$ which
provides a SFR 68\% confidence range of 50--90 M\sol\ yr$^{-1}$,
significantly lower than the range derived from our SED modeling.
However, this UV-to-SFR conversion assumes
constant star-formation over the previous 100 Myr,$^{27}$
whereas our analysis favors substantially younger stellar populations.
Therefore, this conversion will significantly
underestimate the SFR in such galaxies (and caution should be used when
interpreting the SFRs inferred from the UV luminosity that do not
correct for possibly low ages$^{9, }$\cite{bouwens12}).
In the main text, we thus assume the fiducial SFR of 330 M\sol\
yr$^{-1}$, with the caveat here that given uncertainties in modeling
the [O\,{\sc iii}] emission, it may be slightly lower.
In the main text, we discuss the implications of such a high SFR,
assuming that it is due to fueling via gas accretion from the IGM.  
Alternatively, this high SFR could be due to a
merger-induced starburst, which would be detected at its peak SFR with
a $\sim 10-20\%$ probability \cite{neistein08}.  This galaxy does
appear to have a faint companion, though a clumpy
morphology is not necessarily indicative of an ongoing merger\cite{dekel09b}.

As noted in the main text, the best-fit model for the low-redshift
solution has zero [O\,{\sc ii}] emission line flux, inconsistent with
the spectroscopic detection of our emission line, providing further
evidence for our high-redshift solution.  
To see if we could reconcile the photometric
non-detection at $<$ 1 $\mu$m with the detectable emission line flux
if the line were [O\,{\sc ii}], we tried fitting this galaxy with two
populations -- one maximally old (formed at $z =$ 20), and one with an
age and star-formation history which was allowed to vary.  Even including the emission line flux as a
constraint, this fit still preferred a completely passively evolving
model with minimal line emission.

[O\,{\sc ii}] emission at $z =$ 1.78  could be consistent with a
passive population if the galaxy hosted
 an active galactic nucleus (AGN).  This is unlikely as there is no
 {\it Chandra} X-ray source within 30\arcsec.\cite{alexander03}  The {\it
   Chandra} imaging reaches L$_\mathrm{x} =$ 10$^{42}$ erg
 s$^{-1}$ at $z$ = 1.78, sufficient to detect weak AGNs.
To see if an obscured AGN interpretation matches the available data, we
examined the Spitzer/MIPS 24 $\mu$m, {\it Herschel}/PACS 100 and 160
$\mu$m, {\it Herschel}/SPIRE 250, 350 and 500 $\mu$m and the JVLA 1.4 GHz
data \cite{magnelli11,elbaz11,morrison10}.  We found no counterpart to
z8\_GND\_5296 at any of these
wavelengths.  To examine the constraining power of these data, we
compared the spectral energy distribution of the low-redshift obscured
AGN Mrk 231, redshifted to $z =$ 1.78, to the available data.  Such a
galaxy would have been very well detected at all wavelengths.
However, the observed WFC3 and IRAC fluxes for z8\_GND\_5296 are much
fainter than this redshifted template.  Scaling down the template by a
factor of 40$\times$ to match the observed $H$-band flux renders the
$\lambda >$ 24 $\mu$m data unable to constrain this possibility.
However, the observed $H - 3.6 \mu$m color is very inconsistent with
such a template, as we observe this color to be blue, while an
obscured AGN would have a very red $H - 3.6 \mu$m color.  This
inconsistency, combined with the fact that our very deep F814W data
should detect any known $z =$ 1.78 object with our observed WFC3
fluxes, lead us to exclude an obscured AGN as the explanation for this source.

For the high-redshift solution, as shown in Figure 3, the model fitting prefers strong
Ly$\alpha$ emission.  The best-fit model has a Ly$\alpha$ line flux of
4.2 $\times$ 10$^{-17}$ erg s$^{-1}$ cm$^{-2}$, or a factor of
$\sim$15 greater than our observed line flux (likewise, the best-fit
model rest-frame Ly$\alpha$ EW is 120 \AA).  This is certainly due to the way we
treat Ly$\alpha$ in our modeling, where we follow our previous
work$^{13, }$\cite{finkelstein09a}
and assume that half of the line is subject to the IGM 
optical depth at 1215 \AA.  This is analogous to a Gaussian line
symmetric about the resonance wavelength of Ly$\alpha$.  However, as
discussed above, this is rarely the case; in fact, Ly$\alpha$ is
typically observed to be redward of the systemic redshift due to
radiative transfer effects.  However, in these cases, all of the line
flux blueward of 1216 \AA\, as well as many of the photons redward of
resonance (due to the damping wing) are scattered by neutral hydrogen.
Thus, the observed line flux may be severely attenuated from the
intrinsic line flux.\cite{laursen11,dijkstra11} As our stellar population model implies
significant star-formation, it is not surprising that this galaxy may
indeed have a very strong line flux.  The factor of $\sim$15
difference between the best-fit line flux and our observed Ly$\alpha$
line flux could further imply that the Ly$\alpha$ flux of this galaxy is
being severely attenuated, perhaps due to a rising neutral fraction in
the IGM (see the following section).

\begin{figure}
\vspace{-2mm}
\centerline{
\includegraphics[width=0.48\textwidth]{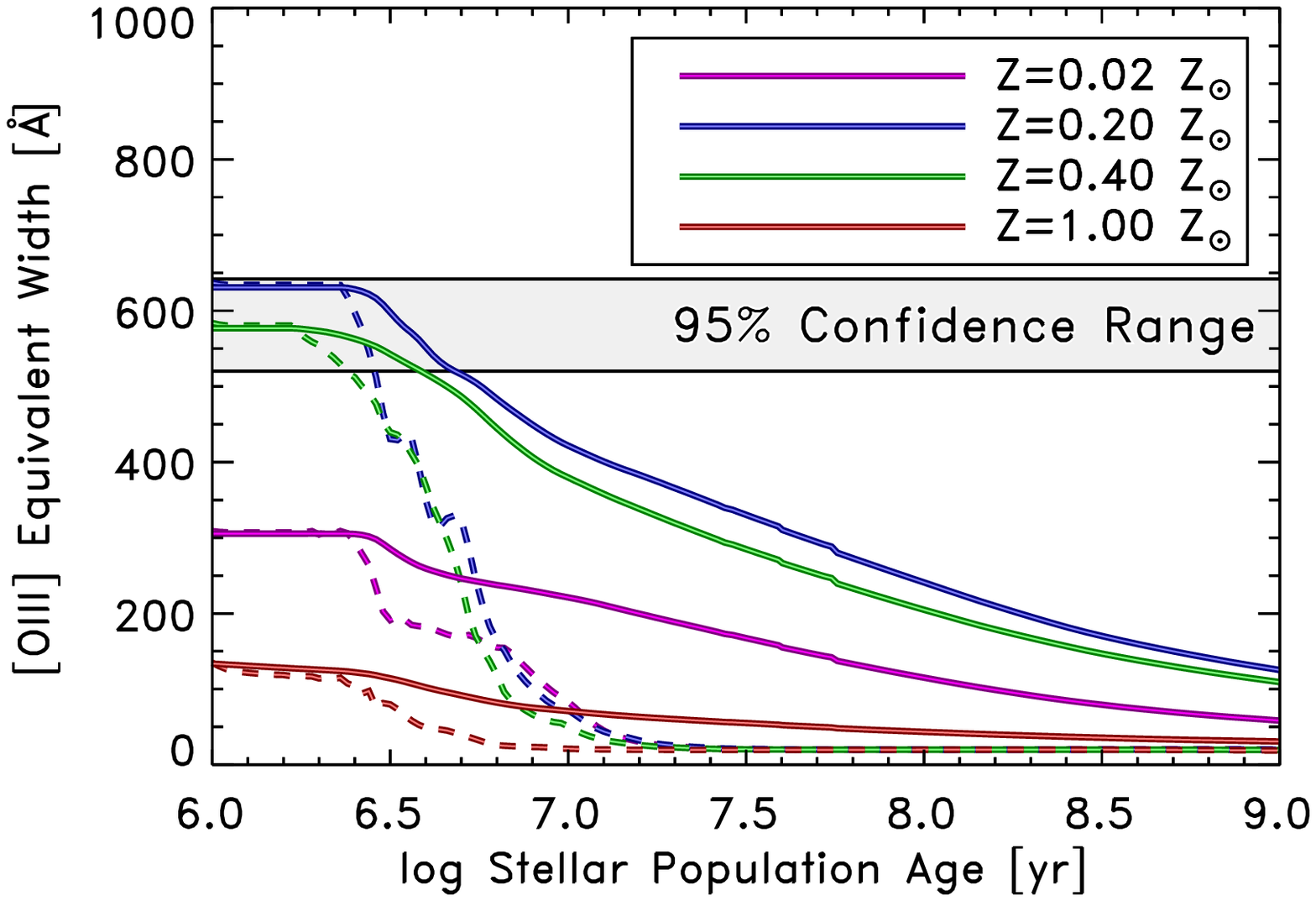}
}
\vspace{-5mm}
\caption{\textbf{[O\,{\sc iii}] EW variation with age and metallicity}. The change of the rest-frame [O\,{\sc
      iii}] $\lambda$5007 EW with stellar population age, for the considered values
    of metallicity.  The solid lines represent a continuous star-formation
    history, while the dashed lines represent an instantaneous burst.
    The 95\% (2$\sigma$) confidence range of our inferred [O\,{\sc
      iii}] EW, 520--640 \AA, is denoted by the gray bar.  At 95\% confidence,
    we can restrict the gas-phase metallicity in this galaxy to be
    sub-solar yet $>$0.02$Z$\sol.
    These results are
    consistent with the stellar metallicity results from the SED
    fitting; which also prefer $Z =$ 0.2--0.4 $Z$\sol.}
\vspace{-5mm}
\end{figure}

\subsection{[O\,{\sc iii}] Emission}
Though typically the small variations of galaxy SEDs with changing
metallicity makes conclusions on the metallicity difficult from
photometry alone, the strong inferred [O\,{\sc iii}] emission in our object
makes at least moderate conclusions possible.
In the main text we discussed how the strong [O\,{\sc iii}] emission
can be used to constrain the metallicity of this galaxy.  Figure S4
shows how the [O\,{\sc iii}] EW varies with age as a function of
stellar population metallicity.  Unfortunately, as we are limited to
the metallicity grid of our chosen stellar population models (which
are not unlike most available models), we cannot make
a precise determination of the limits of the metallicity in this
galaxy.  However, as shown in Figure S4, we can make a few
conclusions.  First, models with solar metallicity cannot come within
a factor of three of creating such high [O\,{\sc iii}] emission, thus
even one of the highest-star-forming galaxies in the distant universe
cannot enrich to $\sim$Solar metallicity by $z \sim$ 7.5.
Secondly, even with a continuous star-formation history, models with $Z =$ 0.02 $Z$\sol\ are still
excluded at $\gg$95\% confidence.  Models with 20 or 40\% Solar metallicity
can reproduce our inferred [O\,{\sc iii}] EW, though with relatively
young ages, consistent with the results from our SED fitting.
Additionally, we also have constraints on the stellar metallicity from
our SED fitting, as we found that all of our 1000 Monte Carlo
simulations preferred a metallicity of either 0.2 or 0.4 $Z$\sol.  The
conservative conclusion from these two results is that $0.02 < Z/Z$\sol$ < 1.0$
at very high confidence, and given the metallicity spacing of our
model grid, 0.2 $< Z/Z$\sol\ $<$ 0.4 is in good agreement with our
measurements.  Further nebular modeling may yield a more precise lower
limit for the metallicity in this system, but the best results will
come from rest-frame optical nebular line spectroscopy with the {\it
  James Webb Space Telescope}.

\begin{figure}
\centerline{
\includegraphics[width=0.48\textwidth]{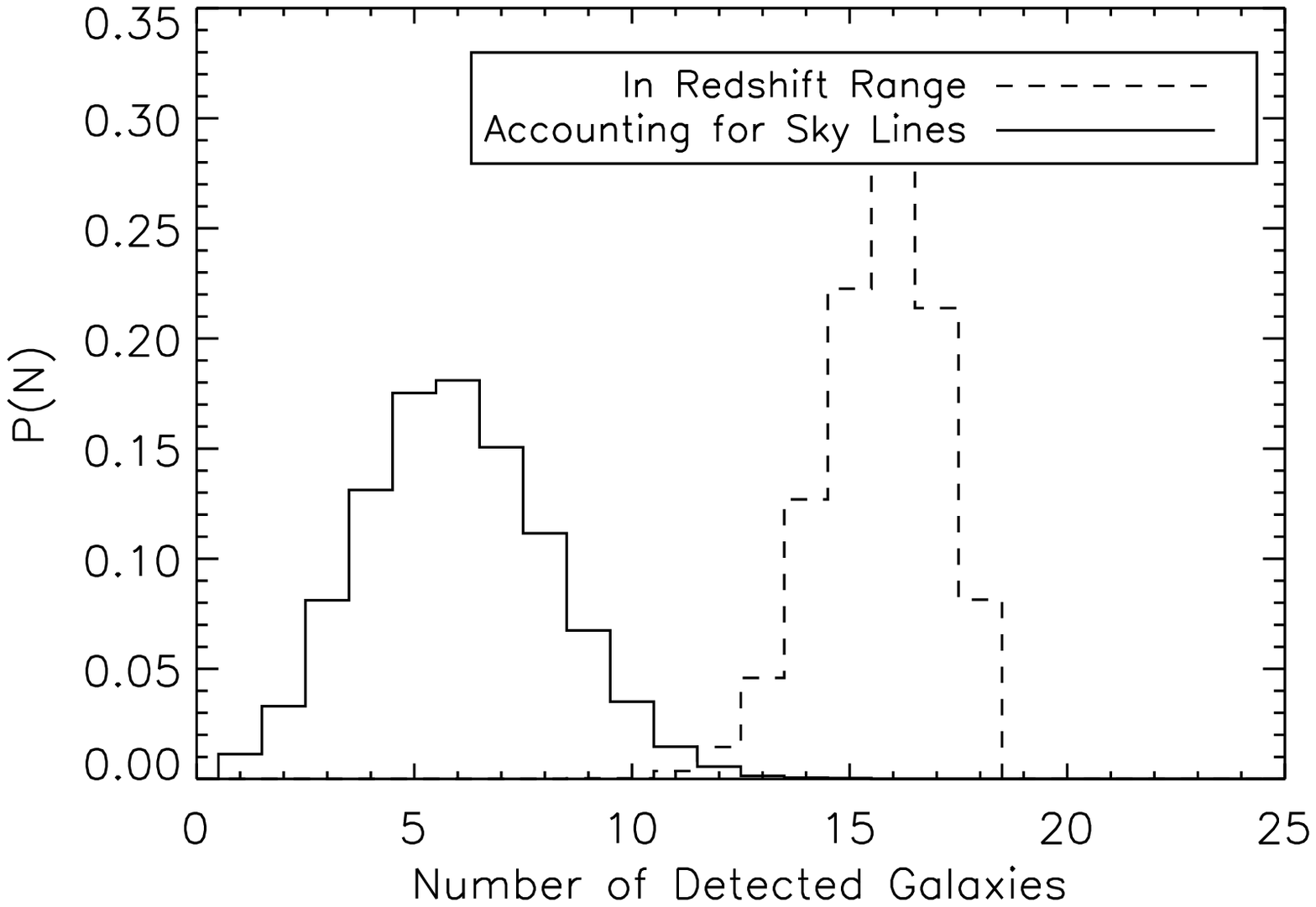}
}
\vspace{-5mm}
\caption{\textbf{EW test}. The results of our Ly$\alpha$ EW
    evolution test, assuming that the Ly$\alpha$ EW distribution at $z
    =$ 7 continues its upward evolution with redshift observed at $z
    =$ 3 -- 6.  The dashed curve shows the expected number of
    detected galaxies in our MOSFIRE data accounting for only the
    spectral range observed.  The solid line shows how this changes if
  we also assume that we will not detect lines which fall on a night
  sky emission line; these sky lines reduce the expected detected
  number by $>$50\%.  Even accounting for this, our simulations
  show that if the EW continues its evolution previously observed at
  $z =$ 3--6 out to $z =$ 7, we would have expected to detect
  Ly$\alpha$ at $>$5$\sigma$ significance from 6 galaxies.  The
  fact that we only detected one such source implies that the
  Ly$\alpha$ EW distribution has evolved at 2.5$\sigma$ significance.}
\vspace{-5mm}
\end{figure}

\section{EW Evolution}
Here we examine the implications of
our lone emission line detection.  We performed a simulation to
predict the number of galaxies we would expect to observe using a
fiducial Ly$\alpha$ EW distribution, with the goal of measuring the
significance at which we could rule out a given distribution. 
For these simulations, we included all
high-redshift candidate galaxies observed on both masks.
 We chose as our EW distribution the predicted $z =$ 7 Ly$\alpha$ EW
distribution from Stark et al.$^{19}$  They use
observations of the evolution of the Ly$\alpha$ EW distribution at 3 $<
z <$ 6 to predict what the distribution would be at $z =$ 7, assuming
the IGM state is unchanged.  
We approximate this distribution as a constant probability from 0 \AA\ $<$
EW $<$ 40 \AA\, then falling off at EW $>$ 40\AA\ as a Gaussian
centered at 40 \AA\ and with FWHM = 60 \AA.
We assigned EWs to our galaxies with a
Monte Carlo approach, in each simulation randomly drawing an EW from
the predicted distribution, and then computing the corresponding
Ly$\alpha$ line flux using the continuum flux of the given galaxy
redward of the line.  In each simulation for each observed candidate
galaxy, we first drew a redshift from the galaxy's redshift
probability distribution function.  If the corresponding Ly$\alpha$ wavelength fell outside the
MOSFIRE $Y$-band spectral range, or if it fell on a sky emission line
(using an extracted sky spectrum to denote the position and extent
of emission lines),
then the galaxy was marked as not detectable.  For all galaxies in
a given simulation which would have Ly$\alpha$ falling in a clean
region of the MOSFIRE $Y$-band, we then compared the simulated
Ly$\alpha$ line flux to the 5$\sigma$ limit of our observations.  If
the line flux was above this value, the candidate galaxy was marked as
detected, otherwise it was left undetected.

For the 5$\sigma$ line flux, we tried two different values.  First, we
assumed our predicted spectroscopic depth of 2.1 $\times$ 10$^{-18}$
erg s$^{-1}$ cm$^{-2}$ (5$\sigma$), from the MOSFIRE exposure time calculator
(this is the value that was used in the main text).  As shown in
Figure S5, this simulation predicts that we should have detected 6.0
$\pm$ 2.2 galaxies.  Out of the 10$^{4}$ simulations run, in only 113
simulations was one or zero galaxies detected at $\geq$ 5$\sigma$, thus
we can rule out this EW distribution at 2.5$\sigma$ significance.  We
note that the consideration of the sky emission lines plays a key
role, as ignoring their presence would have led us to believe that we
should have detected about 10 more galaxies.  As a second test, we
used our flux-calibrated emission line flux of 2.64 $\times$ 10$^{-18}$
erg s$^{-1}$ cm$^{-2}$, at 7.8$\sigma$ significance, to compute an
empirically-derived 5$\sigma$ sensitivity of 1.7  $\times$ 10$^{-18}$
erg s$^{-1}$ cm$^{-2}$.  With this as the limit for detection, we find
that this EW distribution would have predicted 6.0 $\pm$ 2.2 galaxies to
be detected; in this scenario, this EW distribution can be ruled out
at 2.6$\sigma$ significance.  Given the modest uncertainties inherent in our
flux calibration, we consider the first scenario a more
conservative result, though the results are very similar (primarily
because the assumed EW distribution yields predicted line fluxes for most
galaxies brighter than either flux limit).  We will consider the lack of detected emission
lines in more detail in a followup paper (Tilvi et al.\ in prep).